\documentclass[11pt]{article}
\usepackage{epsfig}
\usepackage{latexsym}

\begin{document}
\thispagestyle{empty}
\begin{flushright}
hep-lat/0105023
\end{flushright}
\begin{center}
\vspace*{5mm}
{\LARGE A comprehensive picture of topological excitations 
\vskip2mm 
in finite temperature lattice QCD}
\vskip15mm
{\bf Christof Gattringer${}^\dagger$, Meinulf G\"ockeler, P.E.L.~Rakow, \\
Stefan Schaefer and Andreas Sch\"afer}
\vskip5mm
Institut f\"ur Theoretische Physik \\
Universit\"at Regensburg \\
93040 Regensburg, Germany
\vskip14mm
\begin{abstract}
We study spectra, localization properties and local chirality
of eigenvectors of the lattice Dirac operator. 
We analyze ensembles of quenched SU(3) configurations
on both sides of the QCD phase transition. Our Dirac operator 
is a systematic expansion in path length
of a solution of the Ginsparg-Wilson equation. 
Analyzing the finite volume behavior of our observables and their
scaling with the gauge coupling we come up with a
consistent picture of topological excitations on both sides
of the QCD phase transition. Our results support the interpretation
of the dominant gauge field excitations seen by the Dirac operator 
as a fluid of instanton-like objects. 
\end{abstract}
\vskip3mm
{\sl To appear in Nuclear Physics B.}
\end{center}
\vskip5mm \nopagebreak \begin{flushleft} \rule{2 in}{0.03cm}
\\ {\footnotesize \ 
${}^\dagger$ Supported by the Austrian Academy of Sciences (APART 654).}
\end{flushleft}
\newpage

\setcounter{equation}{0}
\setcounter{page}{1}
\section{Introduction}

The QCD phase transition tightly knits together two of the most intriguing 
features of QCD: When the temperature drops below the critical value 
$T_c$, chiral
symmetry is broken and quarks become confined. State of the art lattice 
results indicate that both these transitions happen at the same $T_c$
(see e.g.~\cite{karsch} for a recent comprehensive review of finite
temperature lattice results).
Lattice studies which compare the changing behavior of
observables on both sides of $T_c$ 
might help to better understand 
the underlying excitations leading to confinement and chiral symmetry 
breaking. 

While different ideas for excitations responsible for confinement 
are still heavily debated, the understanding of chiral symmetry breaking
seems to be in better shape and during the last 20 years a favorite
candidate for the relevant degrees of freedom has emerged. It is now widely
believed that models of instanton-like excitations provide a good
description of chiral symmetry breaking (for extended reviews see 
e.g.~\cite{fintqcd, SchSh98}). 
These models have a wealth of observable signatures 
and here we set out to test these predictions in an ab initio lattice 
calculation and we compare the observables 
on the two sides of the QCD phase transition.

The basic idea of chiral symmetry breaking from instantons 
is based on two arguments: Firstly, the Banks-Casher relation
\cite{BaCa80} (see Eq.~(\ref{baca}) below) relates the chiral condensate
$\langle \overline{\psi} \psi \rangle$ to the density $\rho(\lambda)$
of eigenvalues $\lambda$ of the Dirac operator near to zero. Secondly,
the near zero modes which build up the density of eigenvalues
$\rho(0)$ at small $\lambda$ are believed to come from interacting 
instantons and anti-instantons: As long as an instanton is
well separated from an anti-instanton, the Dirac operator produces 
only zero modes which do not contribute to $\rho(0)$ in the Banks-Casher 
formula. As soon as the instanton and an anti-instanton start to overlap,
the Dirac operator can no longer distinguish them as isolated topological 
objects and instead of two zero modes produces a pair of complex conjugate 
eigenvalues. For small overlap, these two eigenvalues are still very 
close to the origin but move up (down) the imaginary axis as the instanton 
and anti-instanton approach each other.

While earlier studies of instantons on the lattice 
mainly were based on direct filtering of the gauge fields such as 
cooling \cite{cooling}, 
the recent progress for chirally
symmetric lattice fermions based on the Ginsparg-Wilson relation 
\cite{GiWi82} now allows to directly use the spectrum 
and eigenvectors of a chirally
symmetric Dirac operator to test the instanton picture (see 
\cite{ivanenko}-\cite{christ} for studies of spectrum and
eigenvectors of different lattice Dirac operators). 
When using the eigenvectors of the Dirac operator as a probing tool
for instanton-like excitations, 
the basic assumption is that the localization of the 
eigenvector traces the localization of the underlying instanton configuration,
a property which is known for eigenvectors with eigenvalue 0 
of the continuum Dirac operator in an instanton or multi-instanton background.
The scenario for chiral symmetry breaking sketched in the last paragraph 
leads to many characteristic
effects that should be visible in the spectrum and eigenvectors of the 
Dirac operator. In particular the spectral density, the localization of 
eigenvectors and their local chirality give characteristic signatures 
which can be tested in an ab initio calculation on the lattice 
and which might support the instanton picture or rule it out. 

In this article we report on our lattice studies of eigenvectors and 
eigenvalues of the Dirac operator. We use ensembles of SU(3) gauge fields 
in the quenched approximation. Our runs were done on three
different lattice sizes in order to study the volume dependence of 
observables. The gauge couplings were chosen such that we have 
ensembles on both sides of the QCD phase transition. This allows to directly
compare the changing behavior of observables as one changes from the
confined, chirally broken phase to the deconfining, chirally symmetric phase.
Such a direct comparison reveals which degrees of freedom show the most 
drastic changes and thus are relevant excitations for chiral symmetry 
breaking. Our Dirac operator is an approximate solution
of the Ginsparg-Wilson equation based on a systematic expansion 
\cite{Ga00}-\cite{GaHiLa01} and we will refer to it 
as the {\sl chirally improved Dirac operator}. 
It is considerably cheaper to implement 
than e.g.~the overlap operator \cite{overlap} while producing very similar 
results for the relevant eigenvectors and eigenvalues in the physical 
branch \cite{Gaetal01}. This property makes the chirally improved 
Dirac operator ideal for the present study since it allows for very large
statistics which would not have been possible for the overlap operator. 

We simulate finite temperature QCD 
 by working on
a lattice with $L_t < L_s$, using periodic time boundary conditions
for the gauge fields, and antiperiodic for the fermions.
The temperature is given by $T = 1/L_t$.

Let us finally remark on the language convention used in this paper: 
The short and periodic time-direction $L_t$ 
introduces an additional length scale 
which is felt by extended excitations of the gauge field. The solution 
of the SU(3) equations of motion at finite temperature is usually
referred to as {\sl caloron}. Calorons exhibit
a string-like localization in time direction.
However, the transition between
calorons and instantons is continuous, since very small instantons 
hardly feel the finite time extent. Thus we here use the
term {\sl topological excitations} and speak of instantons or calorons 
only when we want to stress particular differences.

The article is organized as follows: In the next section we discuss
the setting of our computations which encompasses a brief presentation of
the chirally improved Dirac operator (Section 2.1), a discussion of the
gauge ensembles (Section 2.2) and the techniques
used for the diagonalization of the Dirac operator (Section 2.3). 
In Section 3 we collect our results for the spectrum of the Dirac operator. 
These results are presented in two parts, Section 3.1 where we discuss 
the density distribution and Section 3.2 where we analyze the fermionic 
definition of the topological charge via the
index theorem. In Section 4 we present our
results for the eigenvectors. In Section 4.1 we give a qualitative discussion
based on some examples followed by a discussion of the localization
properties of the eigenvectors in Section 4.2. In Section 4.3 we present
our results for the local chirality of the eigenvectors. The paper closes with 
a summary in Section 5. In two appendices we give the details of
our Dirac operator (Appendix A) and discuss the fermion zero modes
in a classical caloron solution (Appendix B).

\setcounter{equation}{0}
\section{Preliminaries}

\subsection{The chirally improved Dirac operator}

In the current analysis we use a chirally improved
Dirac operator $D$ which is an approximation to
a solution of the Ginsparg-Wilson equation \cite{GiWi82}. 
$D$ is a systematic expansion 
of a Ginsparg-Wilson Dirac operator
in terms of paths on the lattice. A detailed description of this approach 
was presented in \cite{Ga00,GaHi00,GaHiLa01} and here we only sketch a
few basic steps of the construction.

For about three years it has been understood, that chiral 
symmetry can be implemented through
a Dirac operator $D$ which obeys the Ginsparg-Wilson equation
\cite{GiWi82} (we set the lattice spacing to 1)
\begin{equation}
\gamma_5 \; D \; \; + \; \; D \; \gamma_5 \; \; = \; \; 
D \; \gamma_5 \; D \; .
\label{giwi}
\end{equation}Currently two types of
solutions are known, the so called overlap operator \cite{overlap} 
and perfect actions \cite{fixpd,Haetal00}. 
The overlap operator gives a simple expression for a Dirac operator  
obeying Eq.~(\ref{giwi}), but the involved square 
root of a large matrix,
makes a numerical evaluation quite costly. In two dimensions 
the fixed point Dirac operator has been thoroughly tested in the
Schwinger model \cite{fixpd2d}. In four dimensions the perfect Dirac operator 
has entered the test phase \cite{Halat00}, but 
so far no explicit parametrization has been published.

A third approach to the Ginsparg-Wilson equation is the above mentioned 
systematic expansion of 
a solution of Eq.~(\ref{giwi}). The first step in 
the construction of our chirally improved Dirac 
operator is to write down the most general
Dirac operator on the lattice. This is done by allowing more 
general lattice discretizations of the derivative.
The standard derivative term makes use of nearest neighbors 
only but certainly one can include also
more remote points on the lattice such as next to nearest 
neighbors or diagonal terms etc. Each such term
is characterized by the product of link variables which form the 
gauge transporter connecting the two 
points used in the derivative. The corresponding set of links can be 
viewed as a path on the lattice. The most 
general derivative on the lattice will then include all possible 
paths, each of them with some complex
coefficient. In order to remove the doublers, in addition to the 
derivative terms coming with the Dirac
matrices $\gamma_\mu$ we also have to include terms proportional to 
the unit matrix in Dirac space. To obtain the most
general expression we include all 16 elements $\Gamma_\alpha$ of the 
Clifford algebra, i.e.~we also add tensor, pseudovector
and pseudoscalar terms. To summarize, our Dirac operator is a 
sum over all $\Gamma_\alpha$, each of them 
multiplied with all possible paths on the lattice and each path 
comes with its own coefficient.

The next step is to apply the symmetry transformations: translations, 
rotations, charge conjugation, parity and  
$\gamma_5$-hermiticity which is defined by
\begin{equation}
D \; \gamma_5 \; \; = \; \; \gamma_5 \; D^\dagger \; .
\label{g5herm}
\end{equation}
The property (\ref{g5herm}) can be seen to correspond to the 
properties of $D$ which are used to 
prove CPT in the continuum.  Once these symmetries are implemented 
the coefficients of the
paths in the Dirac operator are restricted. One finds, that groups of 
paths which are related by symmetry 
transformations have to come with the same coefficient, up to possible 
signs. We denote the most
general Dirac operator which obeys the symmetries as:
\begin{eqnarray}
&D& \! \equiv \mbox{1\hspace*{-1.0mm}I} \Big[ s_1 \! <> \; + \;
s_2\! \sum_{l_1} <l_1> 
\; + \; s_3 \!\sum_{l_2 \neq l_1} <l_1,l_2> 
\; + \; s_4 \!\sum_{l_1} <l_1,l_1> \; ... \Big] 
\nonumber
\\ 
&+& \sum_{\mu} \gamma_\mu \sum_{l_1 = \pm \mu} s(l_1) \Big[ \;
v_1\!< l_1 > \; + \; v_2\!\sum_{l_2 \neq \pm \mu} [ <l_1,l_2> + <l_2,l_1> ] 
\nonumber 
\\
& & \hspace*{8cm}
+ \; v_3\!< l_1,l_1> \; ... \; \Big]
\nonumber
\\
&+& \sum_{\mu < \nu} \gamma_\mu \gamma_\nu \sum_{{l_1 = \pm \mu
\atop l_2 = \pm \nu}} s(l_1)\; s(l_2)
\sum_{i,j = 1}^2 \epsilon_{ij} \Big[ \; t_1 <l_i,l_j> \; ... \; \Big]
\nonumber
\\
&+& \!\! \sum_{\mu < \nu < \rho} \gamma_\mu \gamma_\nu \gamma_\rho
\!\! \sum_{{l_1 = \pm \mu, l_2 = \pm \nu \atop l_3 = \pm \rho}} \!\! 
s(l_1)\; s(l_2) \; s(l_3)
\sum_{i,j,k = 1}^3 \epsilon_{ijk} \Big[ \; a_1 <l_i,l_j, l_k> \; ... \; \Big]
\nonumber
\\
&+& \gamma_5 \!\!
\!\! \sum_{{l_1 = \pm 1, l_2 = \pm 2 \atop 
l_3 = \pm 3, l_4 = \pm 4}} \!\! 
s(l_1)\; s(l_2) \; s(l_3) \; s(l_4)
\sum_{i,j,k,n = 1}^4 \epsilon_{ijkn} \Big[ \; p _1 <l_i,l_j, l_k, l_n> 
\; ... \; \Big]\; .
\nonumber \\
\label{gendirac}
\end{eqnarray}
By $\epsilon$ we denote  the totally anti-symmetric tensors with 2,3 and 4
indices. We use the notation $<l_1,l_2 \, ... \, l_n>$ to denote
a path of length $n$ and the $l_i \in \{-4,-3,-2,-1,1,2,3,4\}$ simply
denote the directions of the subsequent links which build up the path. 
$s(l_i)$ is an abbreviation for $sign(l_i)$.
With the particular choice for the generators of the Clifford algebra
used in Eq.~(\ref{gendirac}) (no additional factors of $i$), 
the coefficients $s_i,v_i,t_i,a_i,p_i$ are real. 
The expansion parameter for the Dirac operator in Eq.~(\ref{gendirac}) 
is the length of the path since
the coefficients in front of the paths  decrease in size as the length 
of the corresponding
path increases. A general argument for this behavior can be given and 
it has been 
confirmed numerically for the solution presented in \cite{GaHiLa01}.
We remark, that an equivalent form of $D$ presented in \cite{Haetal00}
is the basis for a 
parametrization of the perfect Dirac operator.

The final step in our construction is to insert the general expression 
for $D$ into the Ginsparg-Wilson
equation. On the left hand side of the Ginsparg-Wilson equation 
(\ref{giwi}) some of the terms
acquire minus signs, depending on the commutator of the corresponding 
$\Gamma_\alpha$ with
$\gamma_5$. On the right hand side an actual multiplication of all the 
terms in D has to be
performed. However, using the above notion of a path, the multiplication 
on the right hand side 
can be formulated in a algebraic way and then can be evaluated 
using computer algebra. 
Once all multiplications are performed one can compare the left and 
right hand sides of 
the Ginsparg-Wilson equation. It is important to note that for an 
arbitrary gauge field different paths,
which correspond to different gauge transporters, are linearly 
independent and can be
viewed as elements of a basis. Thus for the two
sides of Eq.~(\ref{giwi}) to be equal, the coefficients in front 
of the same basis elements on the two sides
have to agree. When comparing the terms on the two sides, 
the result is a set of coupled quadratic
equations for 
the expansion coefficients $s_i,v_i,t_i,a_i,p_i$. This set of 
equations is equivalent to the  
Ginsparg-Wilson equation. After a suitable truncation of 
(\ref{gendirac}) 
to finitely many terms the system can be solved and the 
result is an approximation to a solution
of Eq.~(\ref{giwi}). In addition it is possible
to include a dependence on the inverse gauge coupling $\beta_1$
through an additional constraint for the coefficients. This step allows
to work with less terms in the parametrization. This procedure is similar 
to the tuning of the mass-like shift which is used to optimize the
localization of the overlap operator \cite{overlapimpl}. 
An explicit list of the terms used in 
our parametrization of the Dirac operator and the values of
the coefficients $s_i,v_i,t_i,a_i,p_i$ are given in Appendix A.

After a test of the 2-d chirally improved Dirac operator in the Schwinger 
model with dynamical quarks in
\cite{GaHi00} the construction was outlined for four dimensions 
in \cite{Ga00}. A test of
a Dirac operator based on this approximation was presented 
in \cite{GaHiLa01} and
it was found that the approximation is particularly good in 
the physical part of the spectrum. It was found that 
near the origin the 
deviation of the eigenvalues from the Ginsparg-Wilson circle 
is very small. Since in the present study we 
are interested in the spectrum near the
origin, this makes the chirally improved Dirac operator very 
well suited for the physical
questions analyzed here. It is numerically much less demanding 
than the overlap operator
and has a well ordered spectrum near the origin such that 
spectral densities and the
formation of a gap can be studied. 

Before continuing with the discussion of the gauge action in the 
next subsection, let us discuss
a general property of the eigenvectors of a Dirac operator $D$ which 
obeys the $\gamma_5$-hermiticity
of Eq.~(\ref{g5herm}). Let $\psi$ be an eigenvector of $D$, 
i.e.~$D \psi = \lambda \psi$. Then a 
few lines of algebra show \cite{itoh}
\begin{equation}
\psi^\dagger  \; \gamma_5 \; \psi \; \; = \; \; 0 \; \; , \; \; 
\mbox{ unless } 
\;  \lambda \; \mbox{ is real } \; .
\label{g5sandwich}
\end{equation}
This equation is the basis for the identification of the zero modes. For the 
continuum Dirac operator or also for a $D$ which is
an exact solution of the Ginsparg-Wilson 
equation only zero modes have a non-vanishing matrix element with $\gamma_5$.
Here we are working with an approximate solution of the Ginsparg-Wilson
equation and we do not have exact zero modes. However, eigenvectors
where the corresponding eigenvalue has a non-vanishing 
imaginary part are excluded as candidates for zero modes, since their
matrix elements with $\gamma_5$ vanish exactly. In our approximation the
zero modes show up as eigenvectors with a small real eigenvalue and a
value of $\psi^\dagger \gamma_5 \psi  \sim \pm 0.87$. If one adds more 
terms in 
our approximation then the zero modes would be closer to $0$ 
and the matrix 
element with
$\gamma_5$ closer to $\pm 1$.

\subsection{The gauge ensembles}

The ensembles of gauge configurations for our study were generated with
the L\"uscher-Weisz action \cite{LuWeact,Aletal95}.  
To be specific, we use the improved gauge action as
presented in \cite{Aletal95}. Explicitly, the 
gauge field action reads
\begin{eqnarray}
S[U] & = & \beta_1 \sum_{pl} \frac{1}{3} \; \mbox{Re~Tr} \; [ 1 - U_{pl} ] 
\; + \; 
\beta_2 \sum_{rt} \frac{1}{3} \; \mbox{Re~Tr} \; [ 1 - U_{rt} ] 
\nonumber 
\\
& + & 
\beta_3 \sum_{pg} \frac{1}{3} \; \mbox{Re~Tr} \; [ 1 - U_{pg} ] \; ,
\label{sgauge}
\end{eqnarray}
where the first sum is over all plaquettes, the second sum over all  $2
\times 1$ rectangles and the last sum runs over all parallelograms.
$\beta_1$ is the principal parameter while  $\beta_2$ and $\beta_3$ can
be computed from $\beta_1$  using tadpole improved perturbation theory 
\cite{LeMa93}, giving \cite{Aletal95} 
\begin{equation}
\beta_2 \; = \; - \; \frac{ \beta_1}{ 20 \; u_0^2} \; 
[ 1 + 0.4805 \, \alpha ]
\; \; , \; \;
\beta_3 \; = \; - \; \frac{ \beta_1}{u_0^2} \; 0.03325 \, \alpha
\; .
\end{equation}
with
\begin{equation}
u_0 \; = \; \Big( \frac{1}{3} \mbox{Re~Tr} \langle U_{pl} \rangle 
\Big)^{1/4} \; \; , \; \; \alpha \; = \; - \;
\frac{ \ln \Big(\frac{1}{3} \mbox{Re~Tr} \langle U_{pl} \rangle 
\Big)}{3.06839} \; .
\end{equation}
The couplings $\beta_2, \beta_3$ are 
determined self-consistently from $u_0$  and
$\alpha$ for a given $\beta_1$. In Table~\ref{rundat} we list 
the values of the $\beta_i$ used for our ensembles and our results 
for the expectation value $\langle U_{pl} \rangle/3$ of the plaquette.

\begin{table}[h]
\begin{center}
\begin{tabular}{c|cccc}
$\beta_1 \; \; $ & 8.10 & 8.20 & 8.30 & 8.45 \\
\hline
$\langle U_{pl} \rangle/3 \; \; $ & $+0.6294$ & $+0.6370$ & 
$+0.6432$ & $+0.6521$ \\
$\beta_2 \; \; $ & $-0.5474$ & $-0.5499$ & $-0.5533$ & $-0.5577$ \\
$\beta_3 \; \; $ & $-0.0512$ & $-0.0502$ & $-0.0495$ & $-0.0483$ \\
\end{tabular}
\end{center}
\caption{Parameters for our ensembles of gauge field configurations.
We list the values of the $\beta_i$ and the 
expectation value $\langle U_{pl} \rangle/3$ of the plaquette.
\label{rundat}}
\end{table}
In \cite{GaHiLa01} it was noted that when comparing the expectation value
$\langle U_{pl}\rangle/3$ 
of the plaquette for the L\"uscher-Weisz action with the
corresponding expectation value for the Wilson action one finds that
the former gives rise to a value of $\langle U_{pl} \rangle/3$ 
considerably closer to 1 
(such a comparison makes sense only after matching a physical scale such as 
e.g.~the string tension). Thus the L\"uscher-Weisz action tends to suppress
ultraviolet fluctuations and typically one obtains better results 
for approximate Ginsparg-Wilson fermions \cite{GaHiLa01} as well as
for the overlap Dirac operator \cite{liuetal}. 

For all runs presented here
the time extent of the lattice was held fixed at $L_t = 6$. 
Our values of $\beta_1$ were chosen such that for $\beta_1 = 8.10$ the
system is in the confining phase, while the other three values
($\beta_1 = 8.20, 8.30, 8.45$) are in the deconfined phase. Based 
on a scaling analysis of the spectral gap \cite{gapscale}, we estimate that
the phase transition occurs at $\beta_1 \sim 8.18$. 

In order to set the scale we computed the Sommer parameter
\cite{sommer} $r_0$ for $\beta_1 = 8.10, 8.30$ and
$\beta_1 = 8.45$ on $16^4$ lattices \cite{Gaetal01b}. We give our results for
$r_0/a$ together with the lattice spacing $a$ (assuming $r_0 = 0.5$\,fm)
in Table \ref{statistics}.

\begin{figure}[tb]
\begin{center}
\epsfig{file=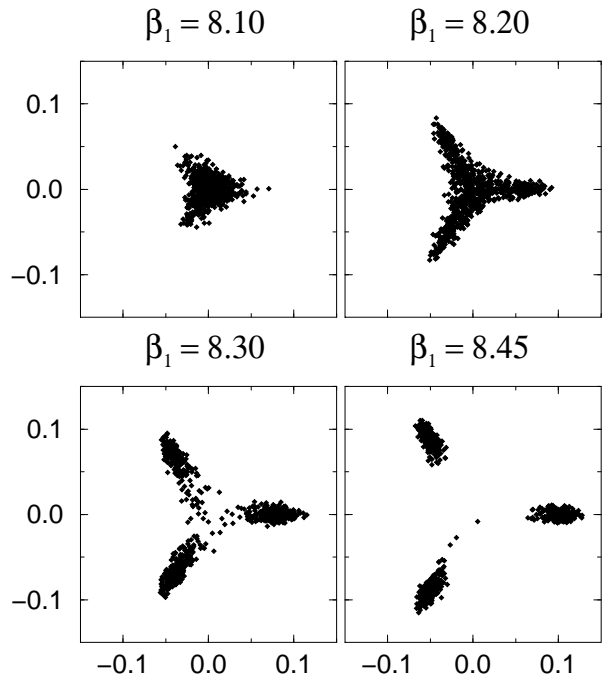,width=8cm,clip}
\caption{The Polyakov loop in the complex plane. 
Each plot shows 800 measurements for lattice size $6\times16^3$.
\label{scatter}}
\end{center}
\end{figure}

In Fig.~\ref{scatter} we show scatter plots of the Polyakov loop 
\cite{poloop}
$\langle L \rangle$
in the complex plane. For each of our samples we display 800
measurements of $\langle L \rangle$ on a $6 \times 16^3$ lattice. 
For $\beta_1 = 8.10$ the measurements of the
Polyakov loop are concentrated at the origin
indicating that the system is in the confined phase. The other three 
plots show the emerging of three disjoint clusters of expectation values 
characteristic for the deconfining phase. The invariance of the
pattern under rotations by $2\pi/3$ is due to the {\bf Z}$_3$ symmetry
of the gauge action. 
The action for fermions does not display this symmetry and
thus fermionic observables behave quite differently
in the sector with real Polyakov loop when compared to the two complex 
sectors \cite{z3pattern}. 
For some of the observables below we will divide our ensembles
of gauge field configurations 
into two sectors according to the phase of the Polyakov loop. We will refer to 
the configurations where the Polyakov loop $\langle L \rangle$ has a phase  
$\sim 0$ as the {\sl real sector}, while configurations with phase 
$\sim \pm 2\pi/3$ will be referred to as the {\sl complex sector}.

Finally we comment on the statistics for our runs. For two 
values of $\beta_1$, one in the confining phase ($\beta_1 = 8.10$), the other
one in the deconfined phase ($\beta_1 = 8.45$), we performed runs at 
three different values of the spatial extent $L_s$ of our lattice 
($L_s = 12, 16, 20$) in order to
study the finite volume behavior of the system. The temporal extent 
$L_t$ was kept fixed at $L_t = 6$. To study the
$\beta_1$-dependence in the deconfined phase we performed two more runs
on the $6\times16^3$ lattice at $\beta_1=8.20$ and $\beta_1=8.30$. 
In order to compensate for the larger fluctuations on the smaller volumes
we analyzed larger samples for these cases. The updates were done with a
mix of Metropolis and over-relaxation steps. We also include
a random {\bf Z}$_3$ rotation in order to 
update this symmetry of the action. Our configurations
are decorelated by roughly two integrated
autocorelations times of the Polyakov loop.
Table \ref{statistics} gives an overview over the statistics for our 
ensembles.

\begin{table}[ht]
\begin{center}
\begin{tabular}{c|cccc}
 & $\beta_1 = 8.10$ & $\beta_1 = 8.20$ & $\beta_1 = 8.30$ & $\beta_1 = 8.45$ \\
\hline
$r_0/a$ & 3.94(16) & - & 4.67(13) & 5.00(5) \\
$a$ & 0.127(5)~fm & - & 0.107(3)~fm & 0.100(1)~fm \\
\hline
$6\times 12^3 \; \;$  & 1200 &   - &   - & 1200 \\
$6\times 16^3 \; \;$  & 800  & 800 & 800 &  800 \\
$6\times 20^3 \; \;$  & 400  &  -  &   - &  400 \\
\end{tabular}
\end{center}
\caption{Sommer parameter $r_0$, lattice spacing $a$ and 
statistics for our gauge field ensembles.
\label{statistics}}
\end{table}

\subsection{Technical remarks}

For the computation of the eigenvalues and 
eigenvectors we used the implicitly
restarted Arnoldi method \cite{arnoldi}.
For each of our gauge field 
configurations we always computed 50 
eigenvalues. The search criterion for the eigenvalues 
was their modulus, i.e.~we 
computed eigenvalues in concentric circles around the 
origin until 50 were found. 

When comparing these eigenvalues for different system 
sizes, one has to take into
account the different density of eigenvalues for 
different volumes. For larger volumes the Dirac operator has
a higher density of eigenvalues 
than on smaller volumes such that when keeping the
number of evaluated eigenvalues fixed at 50, the last eigenvalues found
have larger imaginary parts for smaller lattices when
compared to the results for
larger systems. To give an example, in the chirally broken
phase ($\beta_1 = 8.10$) the largest eigenvalues we found among the 50
computed eigenvalues on the
smallest lattice ($6\times12^3$) typically had imaginary parts
of size $\sim 0.3$ while for the largest lattice ($6\times20^3$)
the largest eigenvalues reached only imaginary parts of $\sim 0.14$
(compare the density plots in the next section). Some of our observables 
are sensitive to these differences due to the technical cut-off of 
50 computed eigenvalues and for them we cut off our data at a
common physical value for all volumes, i.e.~at 
$| \mbox{Im } \lambda | = 0.14$.

\setcounter{equation}{0}
\section{Results for the spectrum}

\subsection{Density distribution of the eigenvalues}
Along with the transition to the deconfined phase, high temperature QCD  
also restores chiral symmetry, leading to a vanishing chiral condensate
$\langle \overline{\psi} \psi \rangle$. Since the chiral condensate is 
related to the density\footnote{Our density $\rho(\lambda)$ is
normalized such that $\rho(\lambda)\, \Delta \lambda$ gives the
number of eigenvalues in the interval $\Delta \lambda$.} 
$\rho(\lambda)$ of eigenvalues $\lambda$ at 
$\lambda = 0$ via the Banks-Casher formula \cite{BaCa80}

\begin{equation}
\langle \overline{\psi} \psi \rangle \; \; = \; \; - \pi \; \rho(0) \; 
V^{-1} ,
\label{baca}
\end{equation}
one expects quite a different behavior of the Dirac spectrum on the
two sides of the transition. In the deconfined, chirally symmetric phase
the density of eigenvalues has to vanish at the origin, up to some 
zero modes due to isolated topological excitations in the gauge fields. 
In the confined, chirally broken phase a non-vanishing density of
eigenvalues extends all the way to $\lambda = 0$.

\begin{figure}[tb]
\begin{center}
\epsfig{file=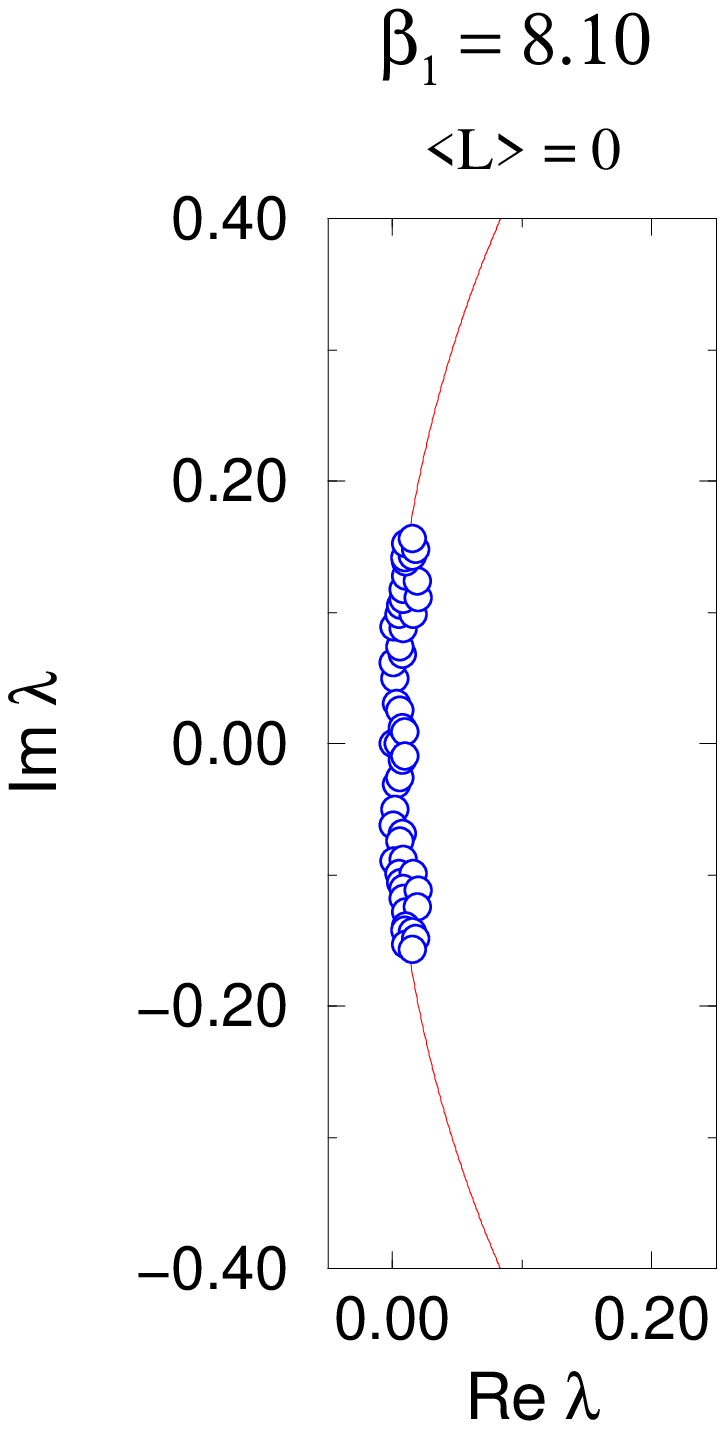,height=8.5cm,clip}
\hspace{1cm}
\epsfig{file=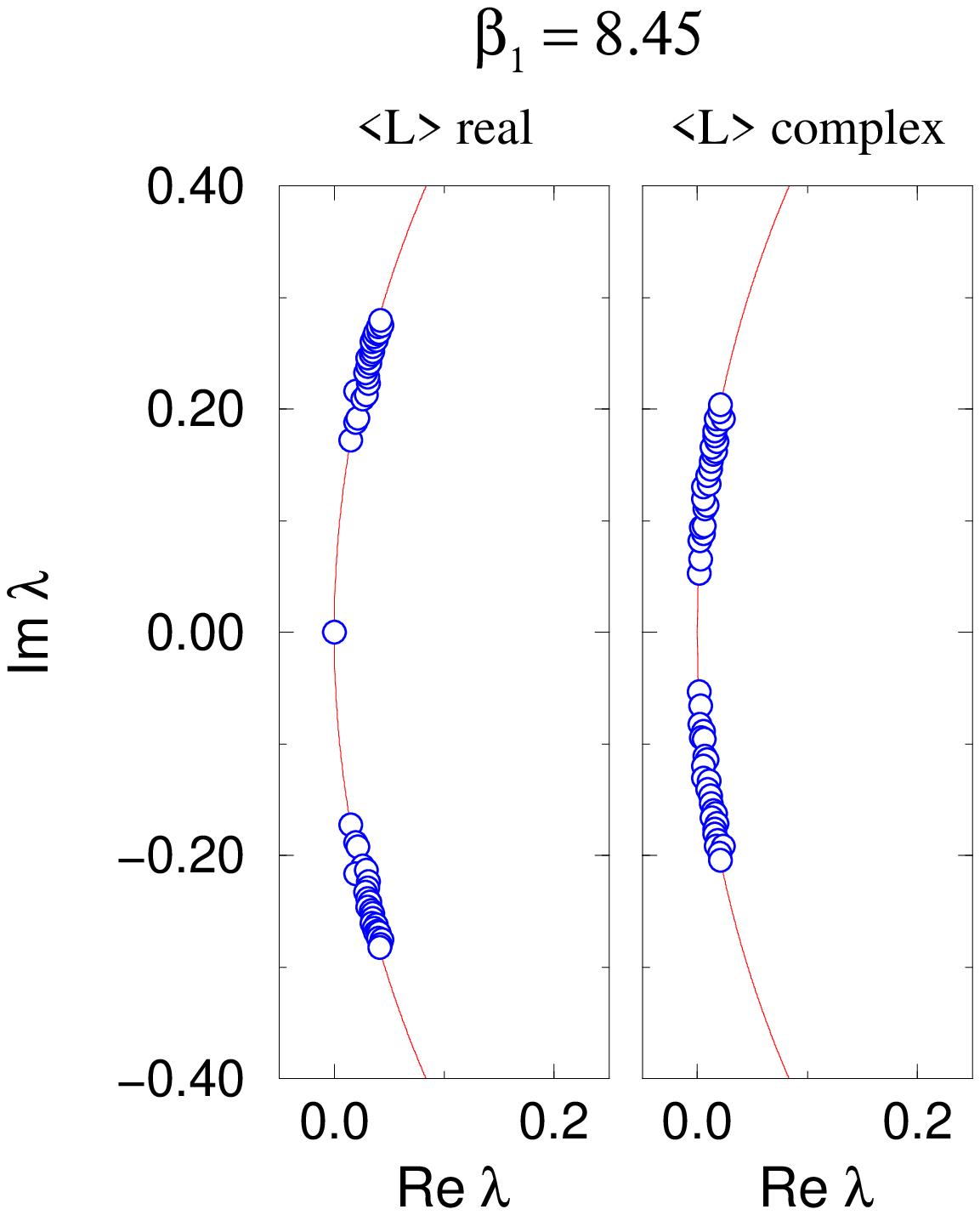,height=8.5cm,clip}
\caption{Examples for spectra of the Dirac operator. We show the
50 eigenvalues closest to the origin, represented by circles. All
spectra are for $6 \times 20^3$ lattices. On the left
hand side we plot the eigenvalues 
$\lambda$ in the complex plane for $\beta_1 = 8.10$
(chirally broken phase). The two plots on the right hand side show spectra
at $\beta_1 = 8.45$ (chirally symmetric phase) for a gauge configuration with
the Polyakov loop in the real sector as well as for a configuration with
complex Polyakov loop. The full curve in the plots indicates the 
Ginsparg-Wilson circle.
\label{spectra}}
\end{center}
\end{figure}

In Fig.~\ref{spectra} we show three characteristic
spectra, each consisting of the 50 eigenvalues closest to the origin
on $6 \times 20^3$ lattices. For the confining, chirally broken phase
($\beta_1 = 8.10$, left hand side of the plot) we find a non-vanishing density
at the origin. When going to the other side of the transition
($\beta_1 = 8.45$, the two plots on the right hand side of Fig.~\ref{spectra})
one finds that the spectrum has developed a gap at the origin. There 
could be some isolated zero modes as e.g.~in the central plot but they
do not contribute to the density $\rho(\lambda)$ in the Banks-Casher
formula (\ref{baca}). As already discussed above, the spectrum behaves
quite differently for gauge configurations in the real and complex sectors 
of the Polyakov loop. For the example with the real Polyakov loop 
(central plot) we find 
a gap roughly twice as big as for the case with a complex Polyakov loop 
shown in the plot at the right hand side. There has been some debate whether 
in the complex sector 
the gap forms at all \cite{z3pattern}, but we find that for
values of $\beta_1$ sufficiently above the critical value the existence
of the gap can be clearly established. A detailed analysis of the volume 
dependence and the scaling of the spectral gap will be presented
elsewhere \cite{gapscale}.

By averaging over all configurations in each sample 
(compare Table \ref{statistics}) one can compute the density 
$\rho(\lambda)$ of
eigenvalues. For the continuum Dirac operator the spectrum is
restricted to the imaginary axis and the definition of
the spectral density is straightforward. For our lattice Dirac 
operator the spectrum is located on the Ginsparg-Wilson circle,
or to be more precise in the vicinity of this circle since here we are
dealing only with an approximate Ginsparg-Wilson operator. In order to
define the spectral density we bin the eigenvalues with respect to 
their imaginary part and count the number of eigenvalues
in each bin. Eigenvalues with vanishing imaginary part, i.e. the
zero modes were left out.
After dividing the count in each bin by the total number
of eigenvalues we obtain the histograms for $\rho(\lambda)$ used in 
Fig.~\ref{valsdens} below. Since we are only interested in the
density $\rho(\lambda)$ in a region of $\lambda$ 
where the real parts of the eigenvalues on
the Ginsparg-Wilson circle are small, binning with respect to the
imaginary parts of $\lambda$ gives a good approximation of
the continuum definition of the spectral density.

\begin{figure}[tb]
\begin{center}
\epsfig{file=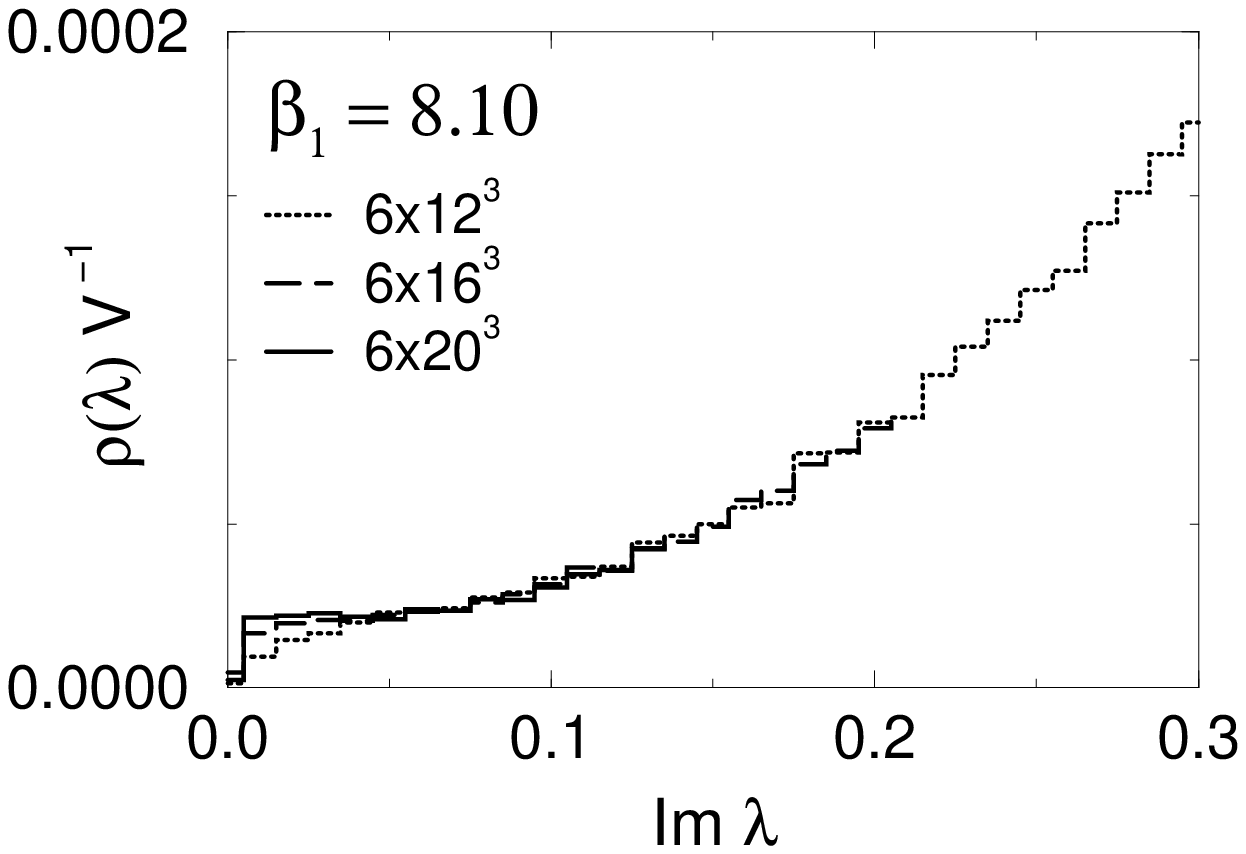,height=5cm,clip}\\
\epsfig{file=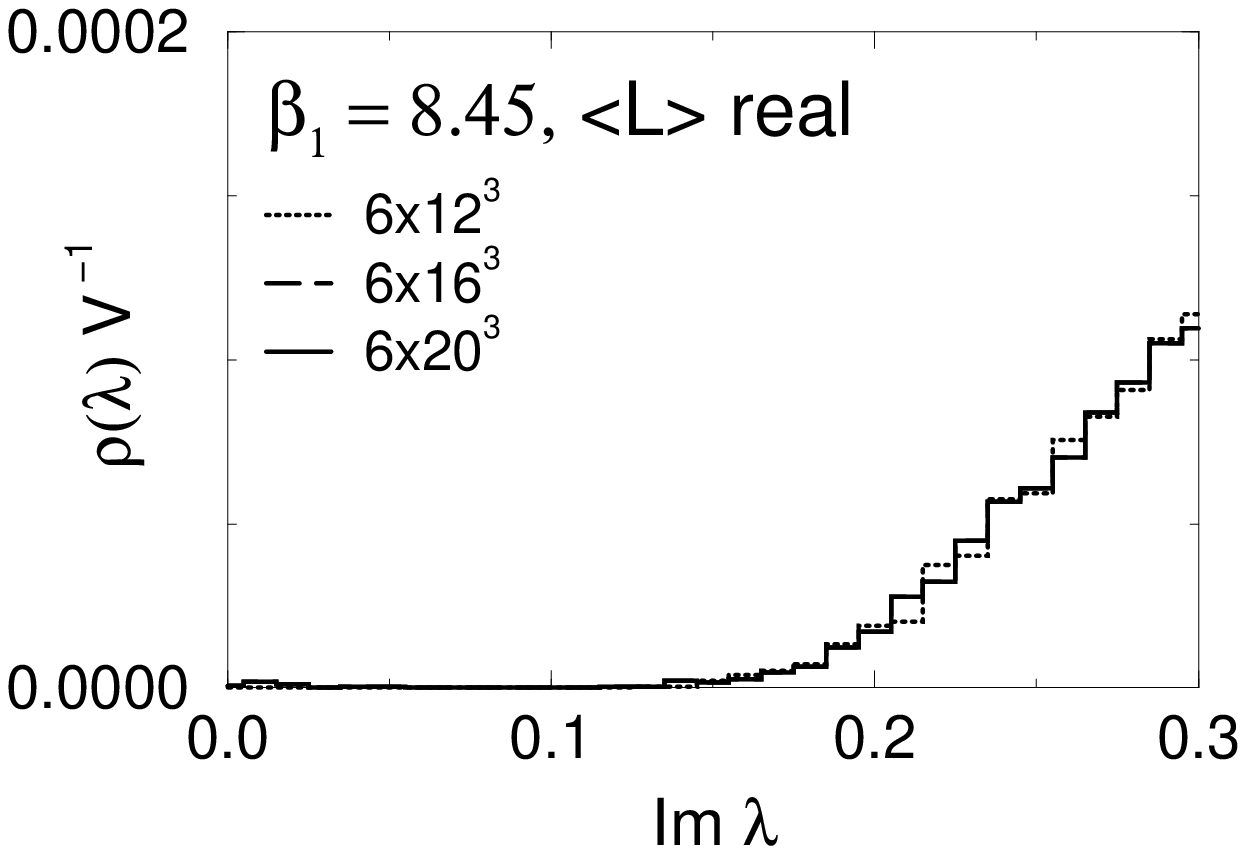,height=5cm,clip} 
\hspace{-4mm}
\epsfig{file=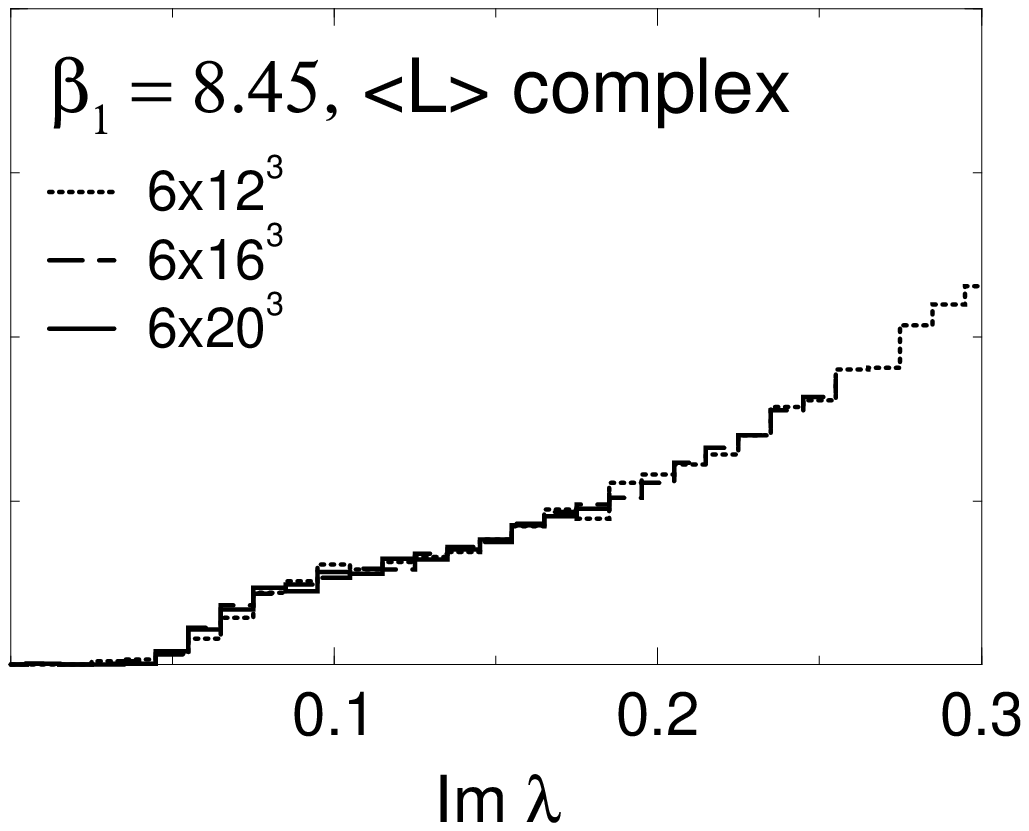,height=5cm,clip}
\caption{Histogram plots for the spectral density $\rho(\lambda)$
normalized with 
the inverse volume $V^{-1}$ as a function of Im $\lambda$. 
\label{valsdens}}
\end{center}
\end{figure}

In Fig.~\ref{valsdens} we show 
(see also \cite{tamas,farch,EdHe00,dametal})
the density of eigenvalues $\lambda$ 
normalized by the inverse volume as a function of Im $\lambda$ 
(we display only the positive half of the curve, 
i.e.~Im $\lambda > 0$). The top plot
gives our results for the confined, chirally broken phase. It is obvious that
after normalization by the inverse volume, the data for different lattice 
sizes essentially fall on a universal curve. Only the edge near the origin  
is more rounded for smaller system size in agreement with
universal random matrix theory predictions \cite{rmth,randommat}. For the
same reason also the curve for the largest system ($6 \times 20^3$) still
shows a drop near the origin. Up to this finite size effect, the density
remains non-zero down to $\lambda = 0$, thus building up the   
chiral condensate according to the Banks-Casher formula (\ref{baca}).
The picture is quite different in the deconfined, chirally symmetric
phase. Again we find that the data for different volumes fall on a 
universal curve, but now
the density drops to zero already at nonvanishing
values of Im $\lambda$. For one of our samples ($6 \times 16^3$)
in the real sector (bottom left plot in Fig.~\ref{valsdens})
we find a very small signal in the vicinity of the origin. We believe that 
this is a statistically insignificant fluctuation 
which does not show up for the larger systems. 
In \cite{EdHe00} a similar event on a $4 \times16^3$ lattice with 
Wilson gauge action at $\beta=5.75$ and overlap fermions was observed
and interpreted as a possible sign for a non-vanishing fermion
condensate in the deconfining phase of quenched QCD.

As already discussed above, the spectral gap 
is considerably smaller for the sector of gauge configurations with
complex Polyakov loop but still is clearly visible (compare also
\cite{gapscale}).

\subsection{Fermionic definition of the topological charge}

In this section we now analyze the zero
modes, which for our Dirac operator correspond
to eigenvalues with vanishing imaginary part. 
These modes come from isolated topological excitations of
the gauge field. In the continuum the 
eigenvectors corresponding to the zero modes can be chosen 
as eigenmodes of $\gamma_5$ with eigenvalues $\pm 1$. The numbers
$n_+, n_-$ of 
these left-, respectively right-handed modes are related to the topological 
charge $Q$ via the Atiyah-Singer index theorem \cite{AtSi71}

\begin{equation}
Q \; \; = \; \; n_+ \; - \; n_- \; .
\label{index}
\end{equation}

A proof of the theorem for the case of a Ginsparg-Wilson lattice Dirac 
operator can be found in \cite{fixpd}. Here we work only with an approximate 
solution of the Ginsparg-Wilson equation and we must give a more
detailed description of our fermionic definition of the topological
charge corresponding to Eq.~(\ref{index}).

\begin{figure}[p]
\begin{center}
\epsfig{file=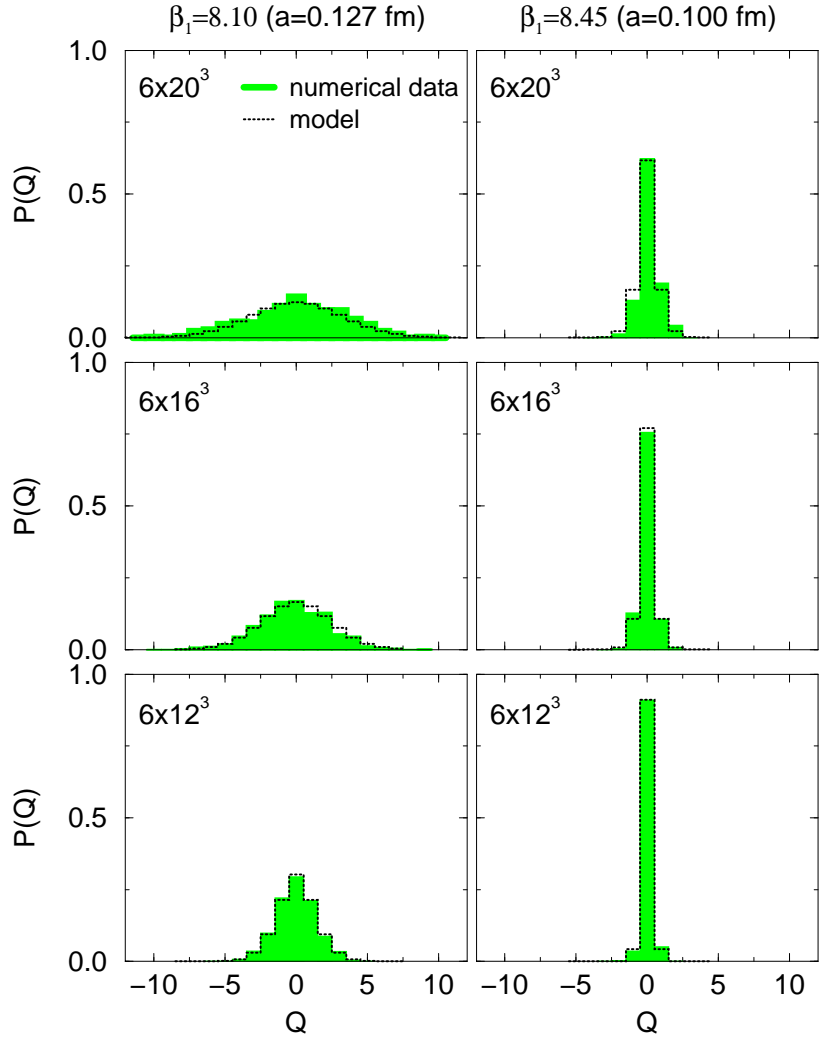,height=14cm,clip}
\caption{Distribution of the topological charge. The filled histograms
show our numerical results and the dotted curve represents the simple
instanton anti-instanton model discussed in the text.
\label{qdist}}
\end{center}
\end{figure}

\begin{figure}[tb]
\begin{center}
\epsfig{file=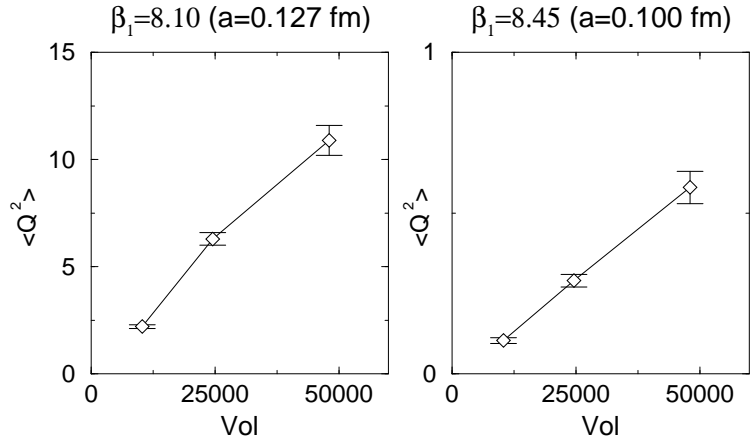,height=6cm,clip}
\caption{Volume dependence of 
$\langle Q^2 \rangle$.
The numerical data are represented by diamonds and we connect the symbols
to guide the eye. Please note the different scale of the vertical
axis in the two plots.
\label{q2dist}}
\end{center}
\end{figure}

There are two main effects of working with only an approximation of a solution
of Eq.~(\ref{giwi}): Firstly, the topological modes do not appear at 
exactly zero, but typically have a small, nonvanishing real part which
increases as the corresponding topological excitation of the gauge field
becomes more like a defect, i.e.~shrinks in size to about a lattice
spacing (compare \cite{Gaetal01}). Secondly, the corresponding eigenvectors 
$\psi$ are not exact eigenstates of $\gamma_5$ but the absolute value of 
$\psi^\dagger \gamma_5 \psi$ is smaller than 1, typically between 
$0.8$ and $0.9$ for our $D$. 
Thus we define $n_+$ ($n_-$) to be the number of eigenvectors
with positive (negative) values of $\psi^\dagger \gamma_5 \psi$
(as noted above $\psi^\dagger \gamma_5 \psi$ vanishes for all
eigenvalues with nonvanishing imaginary part
 (see Eq.~(\ref{g5sandwich})).
In addition we take into account only real eigenvalues 
smaller than 0.15 in order to discard defects. We should remark, that 
when moving this cutoff to either 0.10 or 0.20 our results
as e.g.~shown in 
Fig.~\ref{q2dist} below do not move beyond the error bars. This
implies that for our $D$
defects give only a negligible contribution to the fermionic 
definition of the topological charge. 
  
In Fig.~\ref{qdist} we show our results for the distribution of
the topological charge. The filled histograms show our numerical data
and the dotted curve is the distribution from a simple model of
non-interacting, dilute instantons and  anti-instantons.

The basic assumption is that in a cell of size $\delta V$ the probability
for finding an instanton or an anti-instanton is given by 
$\rho_I \, \delta V / 2$ while the probability for finding an empty cell
is $1 - \rho_I \, \delta V$. Here $\rho_I$ is the density
of the instantons and anti-instantons which 
are assumed to be distributed independently.
After collecting the combinatorial factors one finds that the probability
for having $n_+$ instantons and $n_-$ anti-instantons is given
by a multinomial distribution. This is easily converted to a distribution
$P(Q)$ for the topological charge $Q$ as given 
in Eq.~(\ref{index}) by summing
over $n_+$ and $n_-$ with the constraint 
$\delta(Q - n_+ + n_-)$. When going over to the Fourier transform
it is straightforward to perform the limit $\delta V \rightarrow 0$
which then gives the integrand in Eq.~(\ref{qdistform}) below. 
Inverting the Fourier transform gives the final result

\begin{equation}
P(Q) \; = \; \frac{1}{2\pi} \int_0^{2\pi} dk \; e^{ikQ}
\; e^{- \rho_I V ( 1 - \cos k )} 
\; = \; 
e^{- \rho_I V} \; I_Q(\rho_I V) \; .
\label{qdistform}
\end{equation}
Here $I_Q$ denotes the modified Bessel function of
order $Q$ and $V$ is the
volume of the system. The expectation value $\langle Q^2 \rangle$
is then given by 

\begin{equation}
\langle Q^2 \rangle \; = \; 2 e^{- \rho_I V} \; 
\Big[ \; I_1(\rho_I V) \; + \; 4 \; I_2(\rho_I V) \; + \; 9 
\; I_3(\rho_I V) \; ... \; \Big]
\; = \; \rho_I \, V \; .
\label{topchform}
\end{equation}
Here we used a standard summation formula for the modified Bessel 
functions.
Both results (\ref{qdistform}) and (\ref{topchform}) depend only on
the dimensionless product $\rho_I \, V$. Thus as soon as one 
computes $\langle Q^2 \rangle$ the 
distribution $P(Q)$ is entirely fixed. We remark that in this model
the topological susceptibility $\chi = \langle Q^2 \rangle / V$ is 
simply the density $\rho_I$ of instantons and anti-instantons,
i.e. $\chi = \rho_I$.

 A discussion of more sophisticated dilute instanton models, 
 which take interactions
 into account, can be found in e.g.~\cite{instmodels}.

When comparing our numerical results to 
the theoretical prediction (\ref{qdistform}) in Fig.~\ref{qdist}
we proceeded as follows: $\langle Q^2 \rangle$ was first computed 
for each combination of $\beta_1$ and volume (see our results
below) and then used to evaluate the theoretical distribution $P(Q)$.
Thus the curves shown in Fig.~\ref{qdist} contain no fit. 
We find that on both sides of the transition and for all system sizes
which we analyzed, the numerical data for $P(Q)$
is quite accurately described by the
simple instanton anti-instanton model.

Equation (\ref{topchform}) for $\langle Q^2 \rangle$ also 
makes a simple prediction for the volume dependence of $\langle Q^2 \rangle$. 
In Fig.~\ref{q2dist} we show our results for $\langle Q^2 \rangle$
as a function of the volume. 
For both phases the volume dependence is roughly linear. For the confined,
chirally broken phase ($\beta_1 = 8.10$, left hand side plot) one finds 
a considerably larger value of the topological susceptibility,
i.e.~the slope in the plot of $\langle Q^2 \rangle$ versus $V$  
(note the different scales for the two plots). This
corresponds to a larger density $\rho_I$ in our model. It seems
that in the chirally broken phase the
assumption of diluteness for the instantons and anti-instantons
becomes violated and the topological excitations can no longer be viewed
as non-interacting. This may be the reason for the slight deviation from
a strictly linear behavior of the $\beta_1 = 8.10$ data. For a more accurate 
modeling of the $\beta_1 = 8.10$ data the simple model (\ref{qdistform}),
(\ref{topchform}) would have to be augmented by terms describing an 
interaction of instantons and anti-instantons \cite{SchSh98,instmodels}.

\begin{figure}[ht]
\begin{center}
\epsfig{file=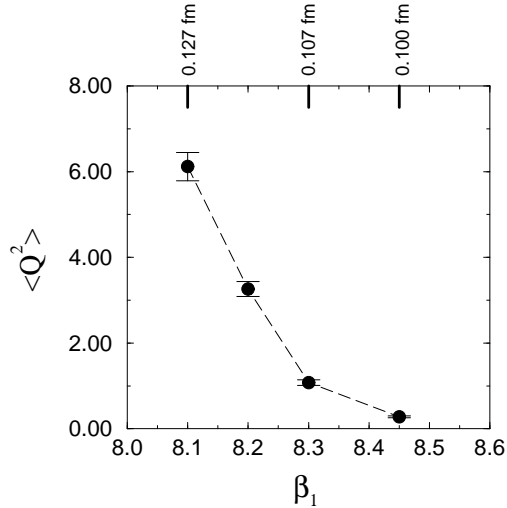,height=7cm,clip}
\caption{$\beta_1$-dependence of 
$\langle Q^2 \rangle$.
The numerical data are represented by filled circles 
and we connect the symbols to guide the eye. The numbers on the
top of the figure give our results for the lattice spacing.
\label{q2vsbeta}}
\end{center}
\end{figure}

Let us now analyze the $\beta_1$-dependence of $\langle Q^2 \rangle$. It
is expected that the chiral susceptibility 
$\chi = \langle Q^2 \rangle / V $ drops considerably as one goes
from the confining, chirally broken to the deconfined, chirally symmetric 
phase. 
In Fig.~\ref{q2vsbeta} we show our results for $\langle Q^2 \rangle$ at
$\beta_1 = 8.10, 8.20, 8.30$ and 
$8.45$. All data are from $6 \times 16^3$ lattices such that 
after a simple rescaling of the vertical axis, Fig.~\ref{q2vsbeta}
is also a plot for the topological susceptibility.
It is obvious that with increasing $\beta_1$ the
values of $\langle Q^2 \rangle$, and thus the values
of $\chi$, drop considerably. 
This implies that isolated
topological excitations are much rarer in the high temperature phase
($\beta_1 = 8.20, 8.30, 8.45$) and that their abundance decreases further as 
one moves away from the critical $\beta_1$. This fits nicely into the above 
discussed picture where the interaction of topological 
excitations gives rise
to the non-vanishing eigenvalue density 
at the origin $\rho(0)$ in the low temperature phase.
Conversely, in the high temperature phase there are simply not enough 
topological excitations to create a sufficient number of
interacting pairs which then 
would build up a non-vanishing eigenvalue density $\rho(0)$. 
Of course, the observable $\langle Q^2 \rangle$ is a measure for the
abundance of zero modes, while a non-vanishing $\rho(0)$ is built up
from the near zero modes. However, it is quite plausible to assume that 
as the net number of isolated topological excitations decreases also the number
of interacting pairs of topological excitations 
goes down. In the next section we 
will confirm this expectation by analyzing 
an observable which is directly sensitive to the number of 
interacting pairs.

\section{Results for the eigenvectors}
\setcounter{equation}{0}
\subsection{A qualitative discussion of some examples}

Before we show our results for the eigenvectors
let us first introduce some notation. Eigenvectors will be denoted 
by $\psi$. They obey the eigenvalue equation
\begin{equation}
D \; \psi \;\;  = \; \; \lambda \; \psi \; .
\end{equation}
The eigenvectors have space, color and 
Dirac indices which we denote by
$x,c$ and $d$ respectively. For each combination of these indices, 
the corresponding entry $\psi(x,c,d)$ 
is a complex number. We define a scalar density $p(x)$ 
and a pseudoscalar density $p_5(x)$
by 
\begin{eqnarray}
p(x) & = & \sum_{c,d} \; \psi(x,c,d)^* \;  \psi(x,c,d) \;
\nonumber \\ 
p_5(x) & = & \sum_{c,d,d^\prime} \; \psi(x,c,d)^* \; 
(\gamma_5)_{d,d^\prime} \; \psi(x,c,d^\prime) \; .
\label{densdef}
\end{eqnarray}
Here ${}^*$ denotes complex conjugation. Note that the 
summation over the color index
makes the densities in Eq.~(\ref{densdef}) gauge invariant. 
Due to the normalization $\psi^\dagger \psi = 1$
of the eigenvectors the scalar density sums up to 1, i.e.
\begin{equation}
\sum_x \; p(x)  \; \; = \; \; 1 \; .
\label{denssum}
\end{equation}

\begin{figure}[p]
\begin{center}
\hspace*{-60mm}
\epsfig{file=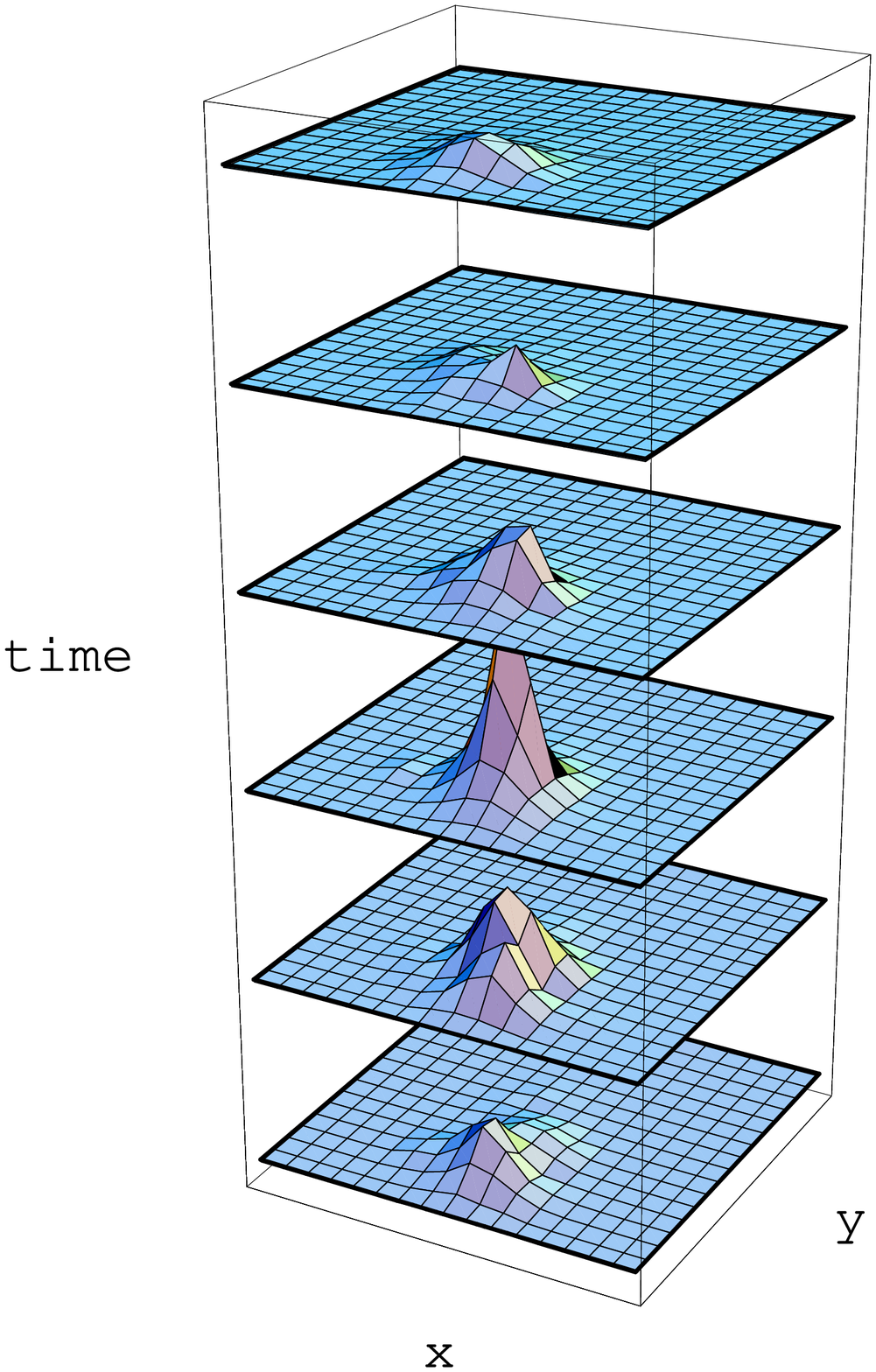,height=11.0cm,clip} \\
\end{center}
\vspace*{-65mm}
\begin{center}
\hspace*{60mm}
\epsfig{file=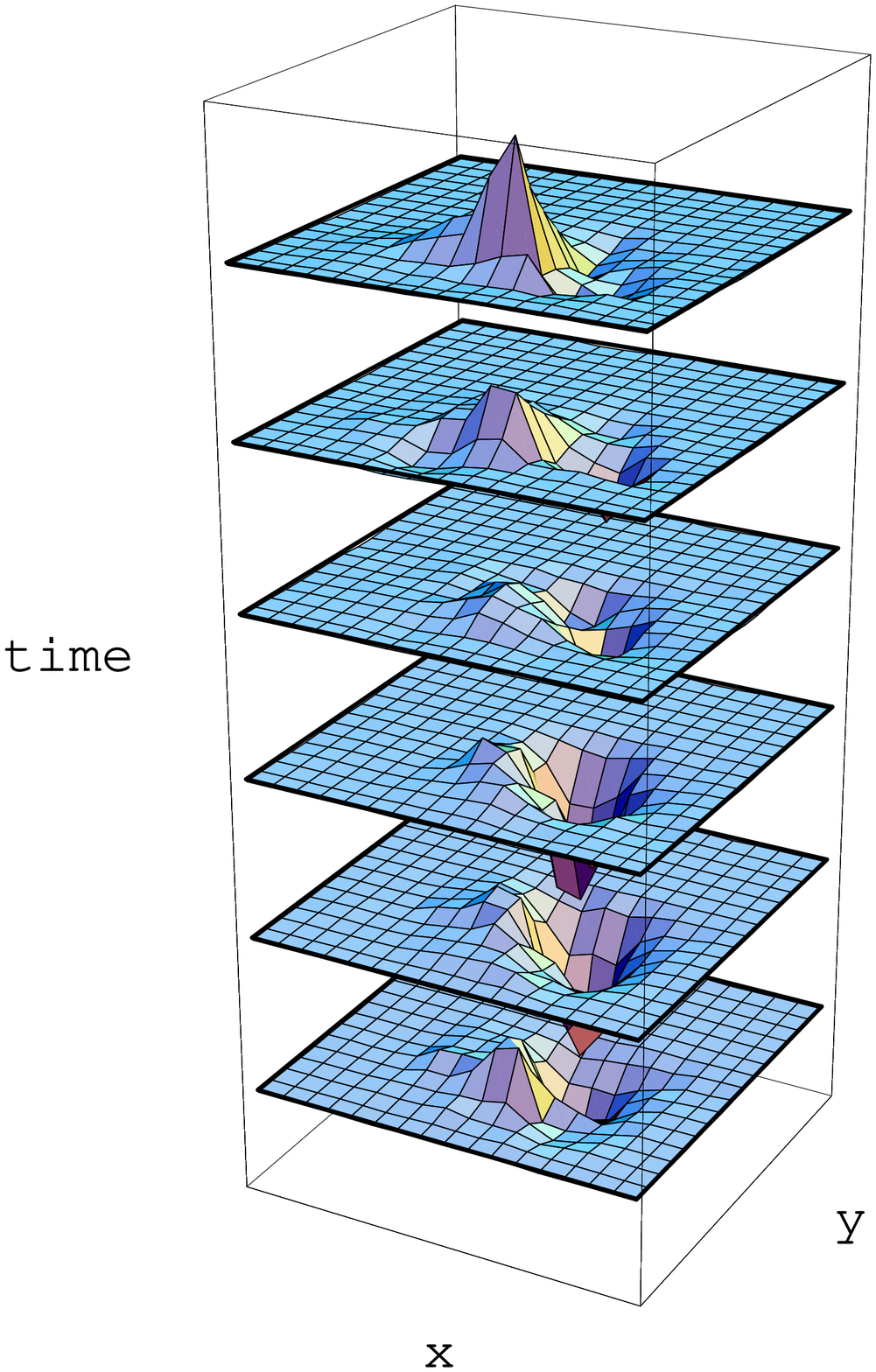,height=11.0cm,clip} 
\caption{The 6 $x$-$y$ time-slices of the 
pseudoscalar density $p_5(x)$ for a zero mode (top left plot) and
a configuration with an interacting caloron anti-caloron pair, a so called
topological molecule (bottom right plot).
The data are from the $\beta_1 = 8.45$ sample on the $6 \times 16^3$
lattice. 
\label{densplots}}
\end{center}
\end{figure}

In Fig.~\ref{densplots} we show $x$-$y$ time-slices for the pseudoscalar 
densities for two
characteristic situations (both eigenvectors
are for configurations from the $\beta_1 = 8.45$ 
ensemble on $6 \times 16^3$ lattices). A similar behavior for the
eigenvectors of the staggered Dirac operator was observed in 
\cite{regensburg}. A study of staggered Dirac operator
 eigenmodes in finite temperature SU(2), \cite{Hands}, also
 found localized modes.  

In the top left plot
we display the six (the temporal extent of our lattice is $L_t=6$) 
$x$-$y$ time-slices of $p_5(x)$ for an eigenvector 
with a single real eigenvalue. As already discussed
above, these modes correspond to isolated topological excitations 
which give rise to a zero mode (in our
setting an eigenvalue with exactly vanishing imaginary part).
Let us discuss some characteristic features of the 
pseudoscalar density 
and compare them to the properties
known for a classical caloron. For a caloron configuration  
the corresponding  
eigenvector (i.e.~the zero mode) roughly traces the action density 
of the gauge configuration, i.e.~$p(x)$
is large where $F_{\mu \nu}^a(x) F_{\mu \nu}^a(x)$ is large. 
Thus $p(x)$ is localized  in space
as is the underlying caloron configuration while it is 
typically  more extended in the
time direction, where it can interact with itself around the short, 
compactified time extension
of the finite temperature lattice.
In addition the eigenvector for a classical caloron is chiral
such that $p_5(x) = \pm \, p(x)$.  
The density $p_5(x)$ displayed in the top left plot
of Fig.~\ref{densplots} shows exactly these features of 
a zero mode corresponding to a caloron
field. The density $p_5(x)$ is localized and positive, i.e.~in this particular
example it corresponds to a caloron which has positive chirality
(negative chirality, i.e.~$p_5(x) < 0$ would correspond to an anti-caloron).

In configurations with several calorons we see several non-degenerate 
real eigenvalues, each having an eigenvector similar to that of the
top left plot of Fig.~\ref{densplots}. Because the modes are not degenerate,
we do not observe strong mixing of the eigenvectors, but instead each of
them is essentially tracing only one of the topological excitations.
For a Dirac operator which is an exact solution of the
Ginsparg-Wilson equation one has exact zero eigenvalues. So in e.g.~a two 
caloron configuration the two zero modes would mix and $p(x)$ and $p_5(x)$
would in general show peaks at the positions of both of the calorons. 

Let us now turn to an eigenvector which has an eigenvalue with 
a relatively small but non-zero imaginary part. 
To be more precise, in the example shown here from the 
$\beta_1 = 8.45$ ensemble this is a mode at the edge of the
spectral gap. According to the caloron picture 
these modes are due to a pair of 
a caloron and an anti-caloron  which approach each other and 
start to interact, such that the
Dirac operator no longer can 
distinguish them as two isolated topological excitations and the 
corresponding two zero modes 
are split into a complex conjugate pair.
We will refer to these configurations of a caloron interacting with 
an anti-caloron as {\sl topological molecule}. 
The dipole structure of such a
topological molecule is nicely illustrated in 
the bottom left plot. The characteristic feature is a positive peak
of $p_5(x)$ (on the $t = 3$ slice) next to a negative peak
(on the $t=6$ slice). 

Finally we remark, that when inspecting 
$p_5(x)$ for an eigenvector
with an eigenvalue in the bulk of the distribution, i.e.~an 
eigenvalue with a large imaginary part one finds that
any trace of a localized topological object 
has been washed out and the pseudoscalar density is dominated by quantum 
fluctuations.

\subsection{Localization properties of the eigenvectors}

In the last section we have presented characteristic examples of 
eigenvectors which display some of the 
features that are expected from models of topological excitations. 
In order to go beyond illustrating
these properties merely by examples, 
in this section we now analyze the sample behavior of
localization properties of the eigenvectors.

A convenient observable for the localization of an eigenvector is
the so-called inverse participation ratio which is widely used
in solid state physics. It is defined by
\begin{equation}
I \; \; = \; \; V \; \sum_x \; p(x)^2 \; .
\label{iprdef}
\end{equation}
Here $V$ denotes the four-volume $L_t \, L_s^3$. From its definition in 
Eq.~(\ref{densdef}) it follows that $p(x) \ge 0$ for all $x$. Taking into
account the normalization Eq.~(\ref{denssum}) one finds that a maximally
localized eigenvector which has support on only a single site $x^\prime$
must have $p(x) = \delta_{x,x^\prime}$. Inserting this into Definition
(\ref{iprdef}) for the inverse participation ratio one finds that a 
maximally localized eigenvector has $I = V$. Conversely, a maximally
spread eigenvector has $p(x) = 1/V$ for all $x$. In this case, the
inverse participation ratio gives a value of $I = 1$. 
Similarly to the scalar inverse participation ratio $I$ of 
Eq.~(\ref{iprdef}) we can also define a measure $I_5$ 
for the localization of the
pseudoscalar density $p_5(x)$ (see Eq.~(\ref{densdef})),  
\begin{equation}
I_5 \; \; = \; \; V \; \sum_x \; p_5(x)^2 \; .
\label{ipr5def}
\end{equation} 
We will refer to $I_5$ as the {\sl pseudoscalar inverse participation 
ratio} (compare \cite{DegHa} for an alternative observable sensitive to
the localization of the pseudoscalar density).

\begin{figure}[tb]
\begin{center}
\epsfig{file=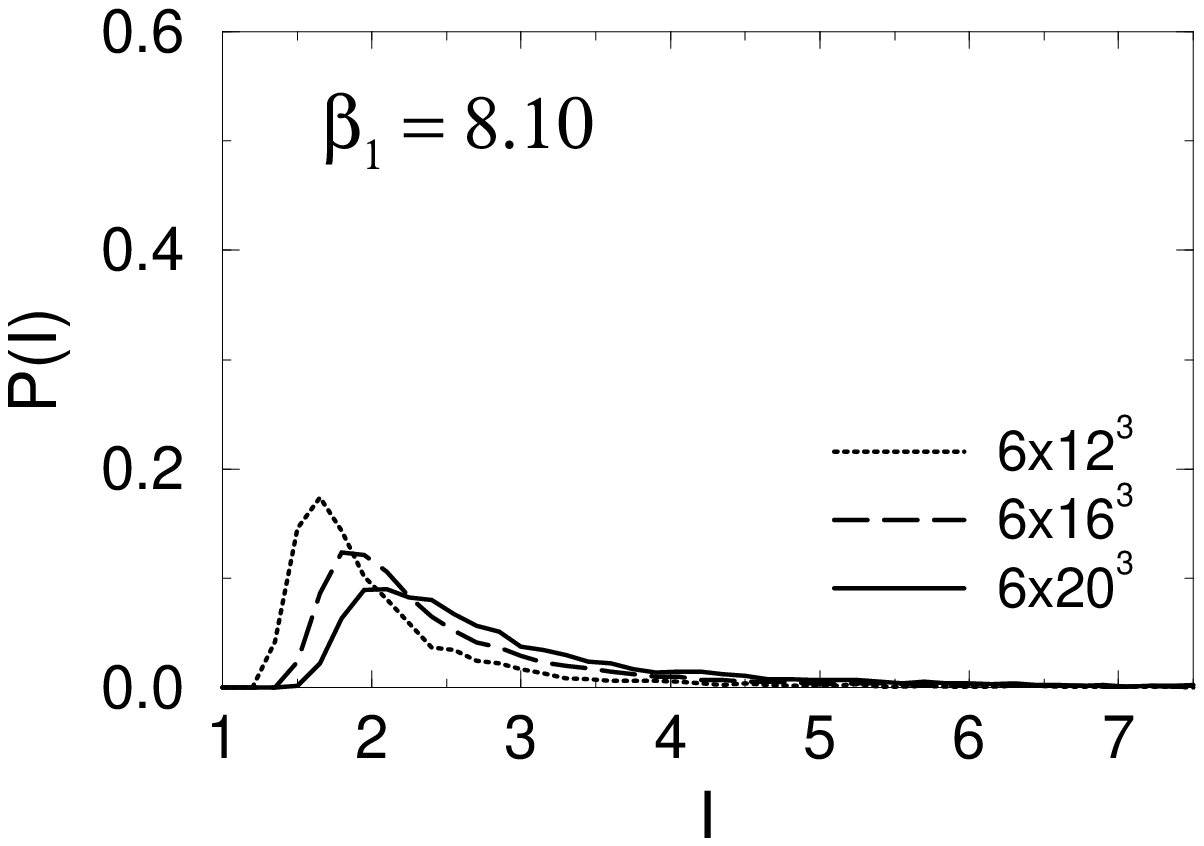,height=4.8cm,clip}\\
\epsfig{file=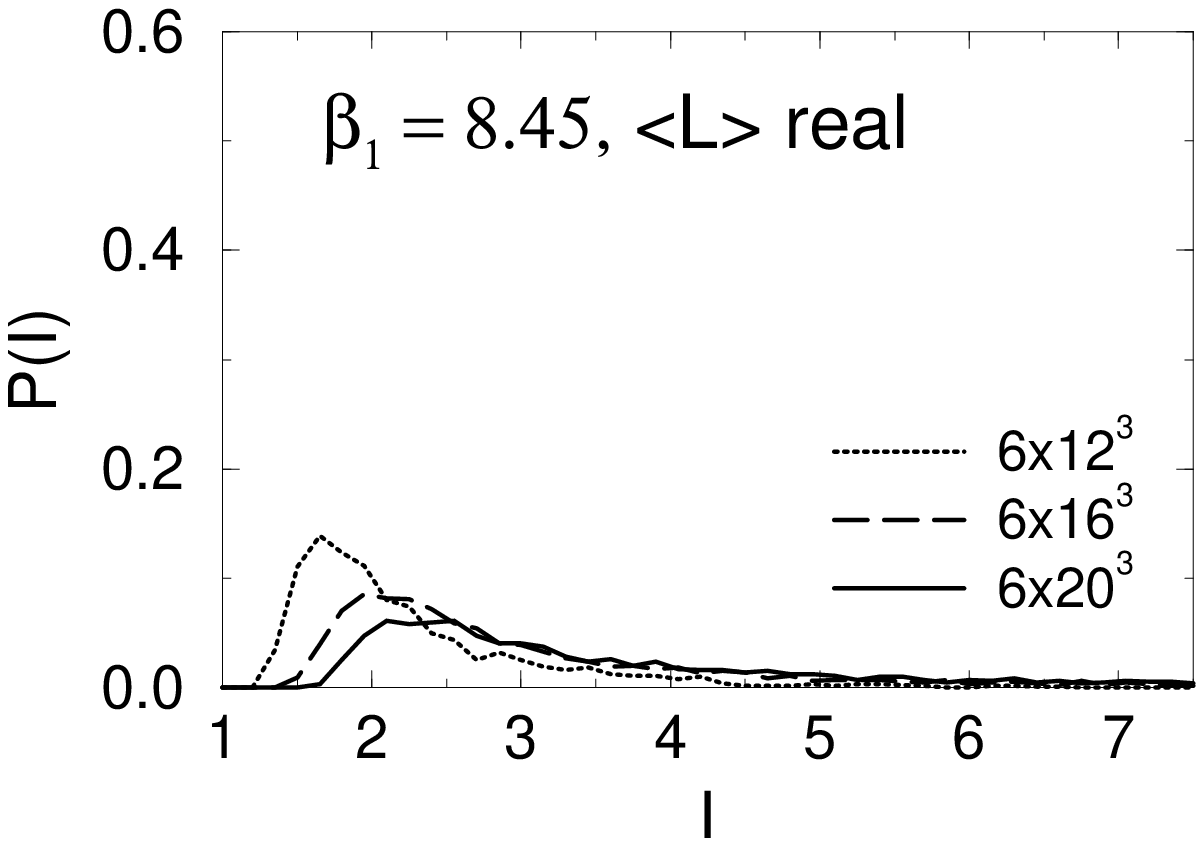,height=4.8cm,clip} 
\hspace{-2mm}
\epsfig{file=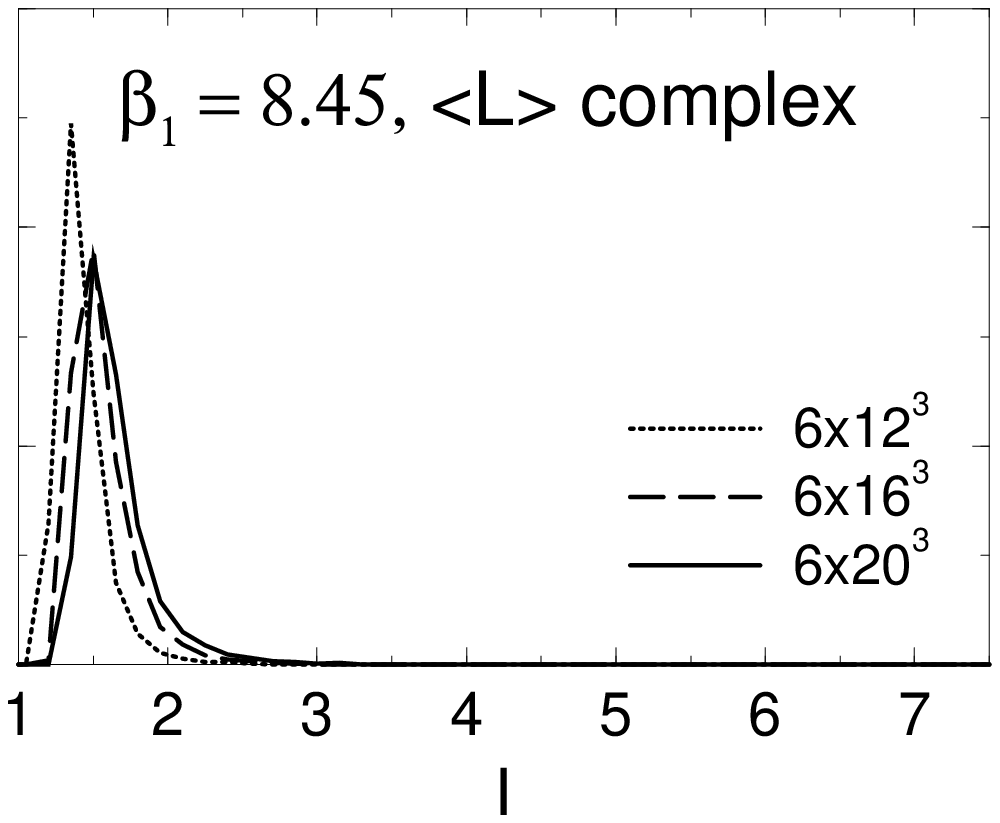,height=4.8cm,clip}
\caption{Distribution of the inverse participation ratio.
\label{iprhist}}
\end{center}
\end{figure}

In Fig.~\ref{iprhist} we show the probability distribution $P(I)$ 
of the inverse participation ratio. We present the results for all
lattice sizes of the $\beta_1 = 8.10$ and $\beta_1 = 8.45$ samples,
where for the latter case we again distinguish between configurations 
with a real Polyakov loop and configurations in the complex sector.
In order to avoid the problem of different cut-off values for
the spectrum when keeping the number of computed
eigenvalues fixed at 50 for all lattice sizes (compare the
discussion in Section 2.3), 
the samples for different volumes were cut off at a common value of the
imaginary part of the eigenvalues.

A general feature for all plots 
in Fig.~\ref{iprhist} is the increasing probability of 
finding localized objects (large values of $I$) as the volume gets
larger. This is consistent with the property of models of 
topological excitations which also grow in number as the volume increases.
When comparing the distribution in the low temperature phase 
($\beta_1 = 8.10$) to
the sector of the high temperature phase ($\beta_1 = 8.45$)
with real Polyakov loop we do not see a large effect. This 
means, that the molecules of calorons which form in the chirally symmetric
phase still are relatively localized objects. 
More significant is the change when one compares 
the plots for the real and complex
sector of the Polyakov loop in the chirally symmetric phase ($\beta_1 = 8.45$,
the two plots at the bottom of Fig.~\ref{iprhist}). In the complex sector 
we find the modes to be much less localized, i.e.~the distribution 
$P(I)$ is much narrower and its peak is shifted to smaller values of 
$I$. An equivalent dependence of the inverse participation ratio
on the  {\bf Z}$_3$ sector
was observed in a study with the staggered Dirac operator in 
\cite{regensburg}.
The dependence of $I$ on the phase of the Polyakov loop can be 
understood from a property of the classical solutions. 
Also for a classical caloron one finds that the eigenvector
becomes more spread out
as one performs a {\bf Z}$_3$ transformation of the gauge field
which rotates the phase of the Polyakov loop (see Appendix B).

\begin{figure}[tb]
\begin{center}
\epsfig{file=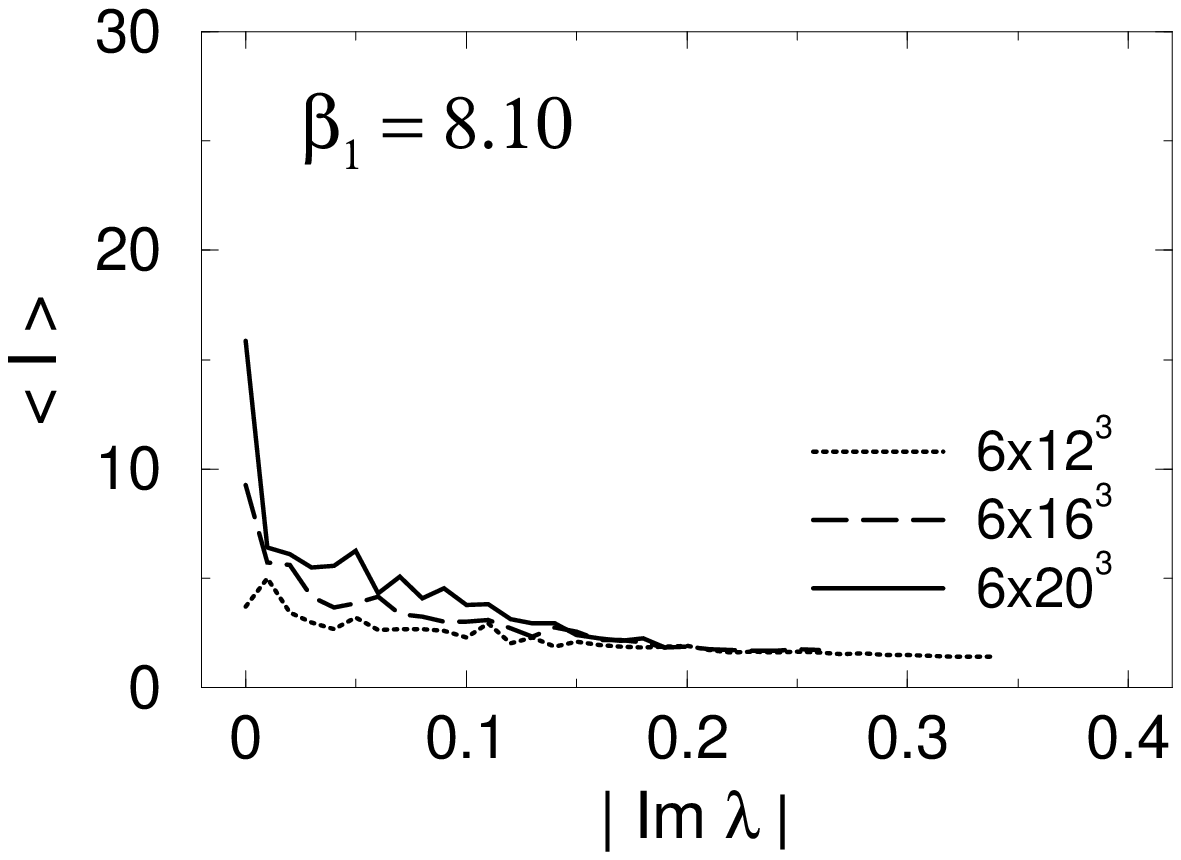,height=4.8cm,clip}\\
\epsfig{file=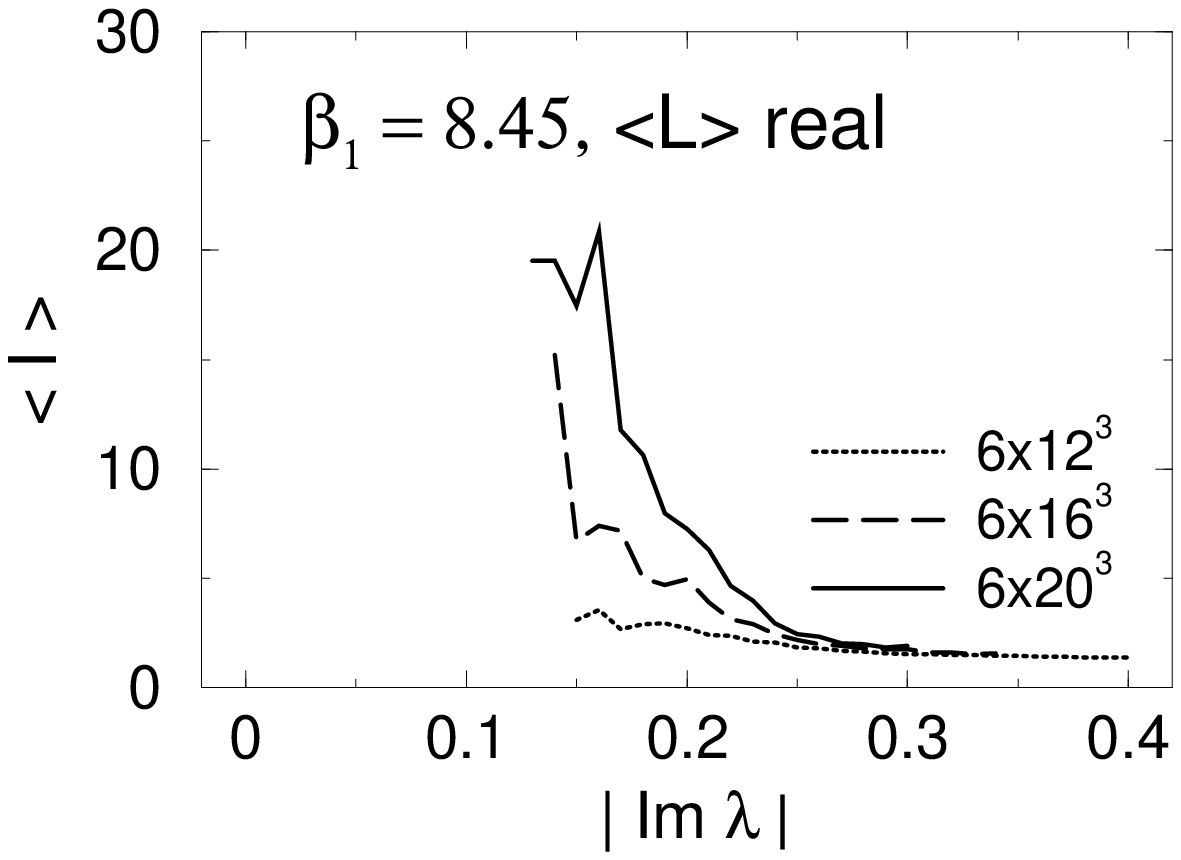,height=4.8cm,clip} 
\hspace{-2mm}
\epsfig{file=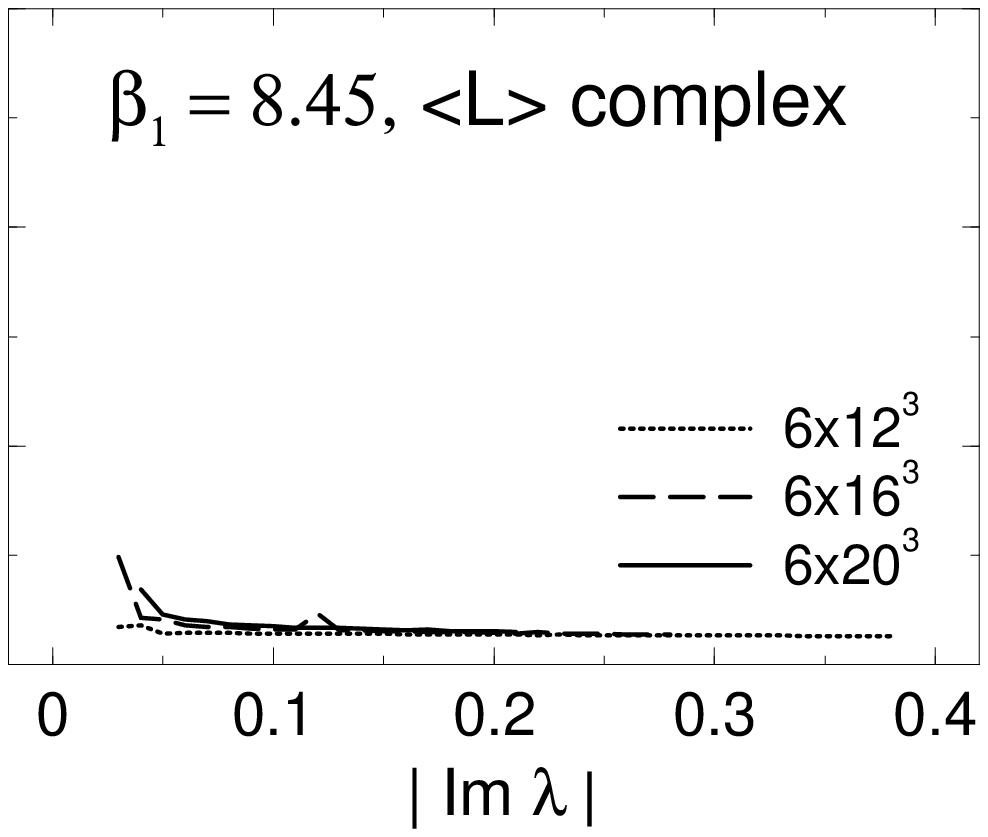,height=4.8cm,clip}
\caption{The inverse participation ratio as a function of 
$|\mbox{Im } \lambda|$.
\label{ivslambda}}
\end{center}
\end{figure}

It is also quite instructive to plot the inverse participation ratio
of an eigenvector against its eigenvalue. In Fig.~\ref{ivslambda} we show 
such a plot. We binned the imaginary parts Im $\lambda$ 
of the eigenvalue spectrum 
and for each bin computed the average size $\langle I \rangle$
of the inverse participation
ratio for the corresponding eigenvectors. Eigenvectors which have a
real eigenvalue, i.e.~the zero modes, were left out.
Let us first discuss the plot
for the chirally broken phase ($\beta_1 = 8.10$,
top plot in Fig.~\ref{ivslambda}). It is obvious that 
for all lattice sizes, the most localized states (they have the largest 
values of $I$) sit near Im $\lambda$ = 0. This behavior can again be 
understood in the context of the picture of a fluid of topological excitations.
As already discussed in the last section, the near zero modes are believed 
to come from pairs of topological excitations which overlap slightly,
i.e.~large topological molecules. As long as the two excitations are  
relatively remote, the corresponding eigenvector is still relatively
localized, i.e.~its $I$ is large.
The closer the two partners come the
more of their localization is washed out and the corresponding eigenvalue
moves away from the origin along the imaginary axis. This behavior is exactly
reflected in the plot where the most localized states are 
near the origin and the localization becomes washed out as one moves 
along the imaginary axis. 

When turning to the chirally symmetric phase ($\beta_1 = 8.45$) 
the behavior is quite similar.
The only difference is the appearance of the gap such that for values of
$|\mbox{Im } \lambda|$ below the edge of the gap we find no
modes at all. Above the edge, however, we again observe that the
most localized states are near the edge and the localization becomes
washed out as one moves further away up or down the imaginary
axis. For the
sector with real Polyakov loop (left hand side plot in the bottom row
of Fig.~\ref{ivslambda})
this behavior is quite pronounced, while due to the 
spreading effect of a {\bf Z}$_3$ rotation discussed
already above and in Appendix B, the signal is much weaker in the 
sector with complex Polyakov loop (right hand side plot in the bottom row).
  
In Section 3 we have seen that $\langle Q^2 \rangle$
drops considerably as one increases $\beta_1$ 
and goes over to the high temperature
phase. $\langle Q^2 \rangle$ is a measure for the net number of
isolated topological excitations. 
In order to study topological molecules,
which do not give rise to zero modes,
we analyze the probability for encountering localized modes. We define
$N(c)$ to be the number of eigenvectors which are 
not zero modes (the corresponding eigenvalue is complex and
$\psi^\dagger \gamma_5 \psi$
vanishes) and have an inverse participation ratio
larger than some cut $c$. Equivalently we define $N_5(c)$ to be the 
number of eigenvectors with a pseudoscalar inverse participation 
ratio $I_5 > c$.

\begin{figure}[ht]
\begin{center}
\epsfig{file=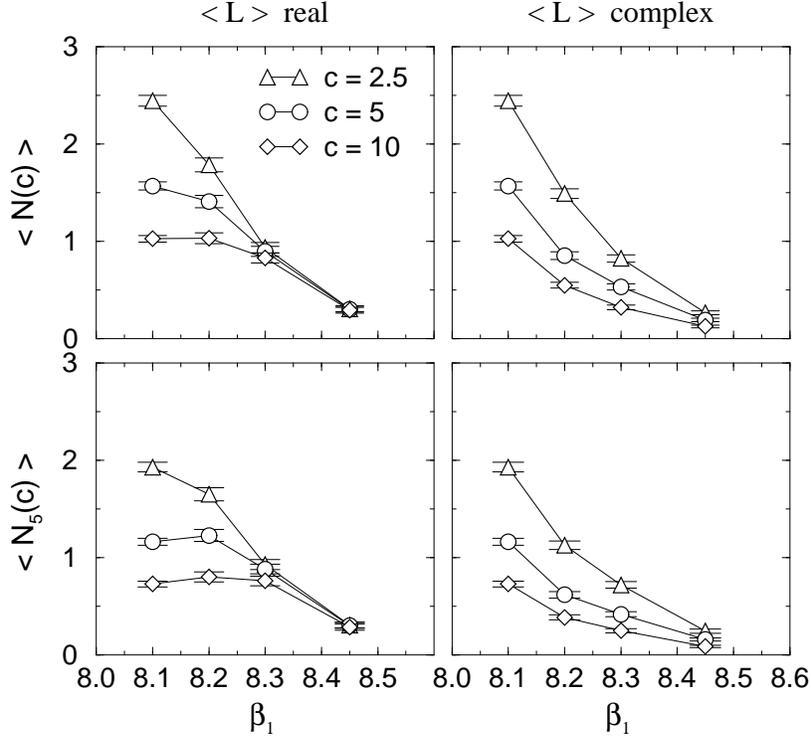,height=10cm,clip}
\caption{Average of $N(c)$ and $N_5(c)$ the numbers of eigenvector
with $I > c$ and $I_5 > c$ respectively, as a function of 
$\beta_1$. Zero modes 
are omitted in the evaluation of $N(c)$ and $N_5(c)$.
We display our results
from $6\times16^3$ lattices for three values of the cut $c$.
\label{i2scal}}
\end{center}
\end{figure}

In Fig.~\ref{i2scal} we show our results for the averages 
$\langle N(c) \rangle$ and $\langle N_5(c) \rangle$
for three different values of $c$. All data were computed on 
lattices of size $6\times16^3$ and the values of $\beta_1$ are 
$\beta_1 = 8.10, 8.20, 8.30, 8.45$. 

It is obvious, that for all three values of $c$ there is a
clear drop of the averages of $N(c)$ and $N_5(c)$ for the numbers of 
eigenvectors with $I > c$ and $I_5 > c$ respectively,
as one increases $\beta_1$. As noted above, zero modes were 
omitted in the evaluation of $N(c)$ and $N_5(c)$ such that the 
observable does not include isolated zero modes and 
is sensitive only to pairs of topological
excitations which, according to the picture of topological
molecules build up the eigenvalue density $\rho(0)$ at the origin.
The drop of $N(c)$ and $N_5(c)$ 
indicates that the abundance of the topological molecules goes down
as $\beta_1$ increases. In the high temperature phase their number  
has become so small that they no
longer create a nonvanishing value of $\rho(0)$ and the
chiral condensate vanishes. 

It is interesting to note that $N(c)$ and $N_5(c)$ drop at essentially
the same rate. Furthermore this holds individually
for both the real and the complex sector
of the Polyakov loop even though the shape of the drop is 
slightly different in the two sectors. This property of a simultaneous 
drop of the average $N(c)$ and $N_5(c)$ further supports 
the above interpretation that the topological molecules simply become
more dilute with increasing $\beta_1$ (which causes the drop of $I$)
but keep their local chirality (which leads to a decrease of $N_5$ at the
same rate as the decrease of $N$).

\subsection{Local chirality of the eigenmodes}

In a recent publication \cite{locchir} Horv\'ath, Isgur, McCune and Thacker
proposed testing the instanton anti-instanton picture
of zero temperature QCD by analyzing
the local chirality of the eigenmodes of the Dirac operator. If the
near zero modes correspond to interacting instantons and anti-instantons,
the corresponding peaks in $p(x)$ (compare Eq.~(\ref{densdef})) should be 
either left or right handed. An example of this behavior is shown in 
Fig.~\ref{densplots}, where left handed peaks show up as positive values of 
$p_5(x)$ and right handed peaks are negative. 
However, using the Wilson Dirac operator
and a relatively small $\beta$ the authors of 
\cite{locchir} did not observe local
chirality of their near zero modes. In the meantime their observable
has been analyzed by different groups using slightly different settings 
\cite{HaDeG01}-\cite{christ}
and all those later publications do indeed see local chirality supporting the 
instanton anti-instanton picture. 

To clarify this point
we also analyzed the local chirality variable of \cite{locchir}. Before
we present our numerical results we briefly repeat the definition of
the local chirality variable of \cite{locchir}.
Similar to the densities $p(x)$ and $p_5(x)$ 
of Eq.~(\ref{densdef}) we can
now define local densities $p_+(x)$ and $p_-(x)$
with positive and negative chirality 

\begin{equation}
p_\pm(x) = \sum_{c,d,d^\prime} \; \psi(x,c,d)^* \; 
(P_\pm)_{d,d^\prime} \; \psi(x,c,d^\prime) \; ,
\end{equation}
where $P_\pm$ are the projectors onto positive and negative 
chirality $P_\pm = (1 \pm \gamma_5)/2$. It is now interesting 
to analyze locally for each lattice point $x$ the ratio
$p_+(x)/p_-(x)$. For a classical instanton in the continuum, 
the corresponding
zero mode $\psi(x)$ has positive chirality and the density $p_+(x)$
is positive for all $x$ while $p_-(x)$ vanishes everywhere. Thus
the ratio of the two always gives $\infty$. For an anti-instanton the 
roles of $p_+$ and $p_-$ are exchanged and the ratio is always 0. 
When now analyzing the eigenvectors for an interacting instanton 
anti-instanton pair this property should still hold approximately
near the peaks. The ratio $p_+(x)/p_-(x)$ is expected to be
large for all $x$ near the instanton peak
of the pair and small for all $x$ near the anti-instanton peak. 
In a final step Horv\'ath et al.~map the two extreme values $\infty$ and 
$0$ of the ratio $p_+(x)/p_-(x)$ onto the two values $+1, -1$ 
using the inverse tangent. One ends up with the local chirality variable
$X(x)$ defined as 

\begin{equation}
X(x) \; = \; \frac{4}{\pi} \; 
\tan^{-1} \left( \sqrt{\frac{ p_+(x) }{ p_-(x) }} \right) \; - 
\; 1 \; .
\end{equation}
When the eigenvectors 
for the near zero eigenvalues correspond to instanton anti-instanton
pairs then the values of $X(x)$ should cluster near $+1$ and $-1$ 
when one chooses lattice points $x$ near the peaks of the
scalar density $p(x)$. One can choose different values for the 
percentage of points $x$ shown and here we will present results
for cuts of 1\%, 6.25\% and 12.5\%. This means that we 
average over those 1\% (6.25\%, 12.5\%) of all lattice points 
which support the highest peaks of $p(x)$. 
The smallest cut-off of 1\% will give 
the best signature since only the highest peaks which are not so
much affected by quantum fluctuations do contribute. On the other
hand such a small cut-off may not be very conclusive, since typically 
instanton models have a packing fraction of
instantons considerably larger than 1\% (see \cite{SchSh98,instmodels}).

\begin{figure}[p]
\begin{center}
\epsfig{file=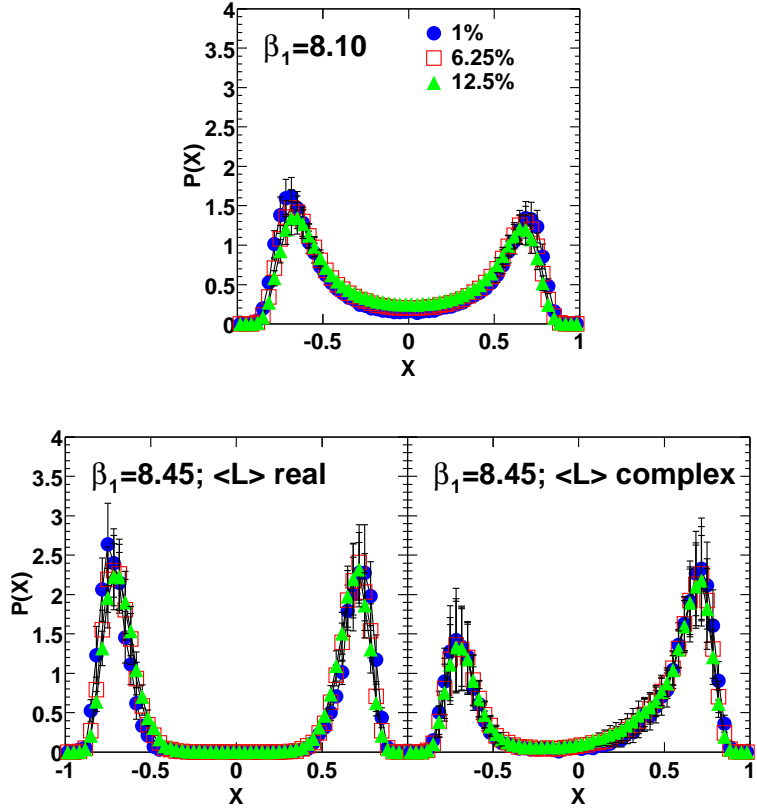,width=10cm,clip} 
\caption{Local chirality for the zero modes. We present data
for both sides of the phase transition. For the chirally symmetric
phase, we distinguish between the real and the complex sector 
of the Polyakov loop. We use different values for the cut-off
on the percentage of supporting lattice points: 1\%, represented by 
filled circles, 6.25\%, open squares and 12.5\%, filled triangles.
The data were computed on the $6 \times 20^3$ lattice.
\label{lchirzero}}
\end{center}
\end{figure}

We would like to remark that in \cite{Wuppertal} a modification
of the local chirality variable $X$ was presented, which uses 
a bi-orthogonal system to construct $p_\pm(x)$. This considerably
improves the signal for the Wilson Dirac operator. We tested 
the version proposed in \cite{Wuppertal} and found only a minor improvement
of our results. This is due to the fact that we use an 
approximation of a Ginsparg-Wilson Dirac operator. A solution 
of the Ginsparg-Wilson equation is a normal operator which can be diagonalized 
with a unitary transformation and the bi-orthogonal system then reduces
to the right eigenvectors used here.

We start the discussion of the local chirality with the zero modes.
In Fig.~\ref{lchirzero} we show
our results for the distribution of $X$ on both sides of the phase transition.
The data for the chirally broken phase ($\beta_1 = 8.10$) was computed
from 55 gauge configurations. For the chirally symmetric phase 
($\beta_1 = 8.45$) we also analyzed 55 gauge configurations but
here we distinguish between configurations with real and complex 
Polyakov loop, such that the sample size for each of the sectors
individually is only 
one third, respectively two thirds of the overall size of 55. In addition,
zero modes are much rarer for the $\beta_1 = 8.45$ sample 
(compare e.g.~Fig.~\ref{qdist}) and the two effects combined 
account for the larger error bars in the $\beta_1 = 8.45$ data.
We also attribute the slight asymmetry observed in the data for the 
complex sector of the Polyakov loop to a fluctuation in the 
relatively small statistics for
this plot which seems to contain more left-handed than right-handed modes
(compare also Fig.~\ref{qdist}).
All data we show were computed for the largest volume $6\times20^3$.
We remark that our data for $P(X)$ are normalized such that the area under 
the curve is 1.

All three plots in Fig.~\ref{lchirzero} show a clear dip of $P(X)$
near $X = 0$. Furthermore the shape of $P(X)$ is almost independent of the 
percentage of lattice points chosen. This shows that zero modes of 
the Dirac operator are locally chiral up to at least one
eighth of the total volume. It is remarkable that the local chirality
signal for the zero modes in the chirally broken phase ($\beta_1 = 8.10$)
is less pronounced (the two peaks are smaller and wider) than for the zero
modes in the chirally symmetric phase. This might be due to the fact
that the smaller value of $\beta_1$ allows for fluctuations on smaller 
scales (measured by the number of lattice points) which spoil the
signal for local chirality. These UV fluctuations 
become more suppressed as one increases $\beta_1$.
Still we can summarize: We find a clear signal for local chirality of the zero 
modes on both sides of the phase transition.

\begin{figure}[p]
\begin{center}
\epsfig{file=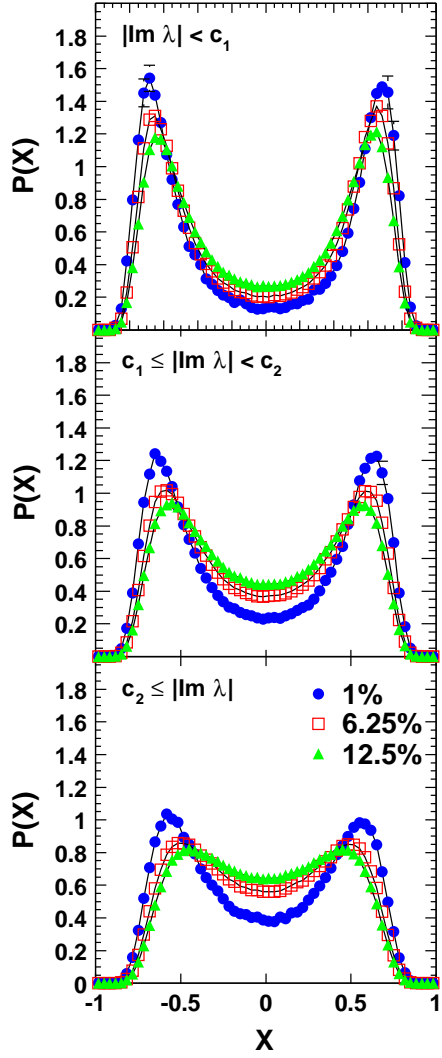,height=14cm,clip} 
\caption{Local chirality for non zero modes in the chirally broken phase.
We bin the eigenvectors with respect to the imaginary parts of
the corresponding eigenvalues such that the first bin
($|\mbox{Im } \lambda | < c_1$, top plot) shows the very near zero modes,
while the other two bins ($c_1 \leq | \mbox{Im } \lambda | < c_2$, middle plot
and $c_2 \leq | \mbox{Im } \lambda |$, bottom plot) show the 
local chirality for eigenvectors with eigenvalues 
higher up in the spectrum. The choice of the thresholds $c_i$ is
discussed in the text.
\label{lchir810}}
\end{center}
\end{figure}

\begin{figure}[p]
\begin{center}
\epsfig{file=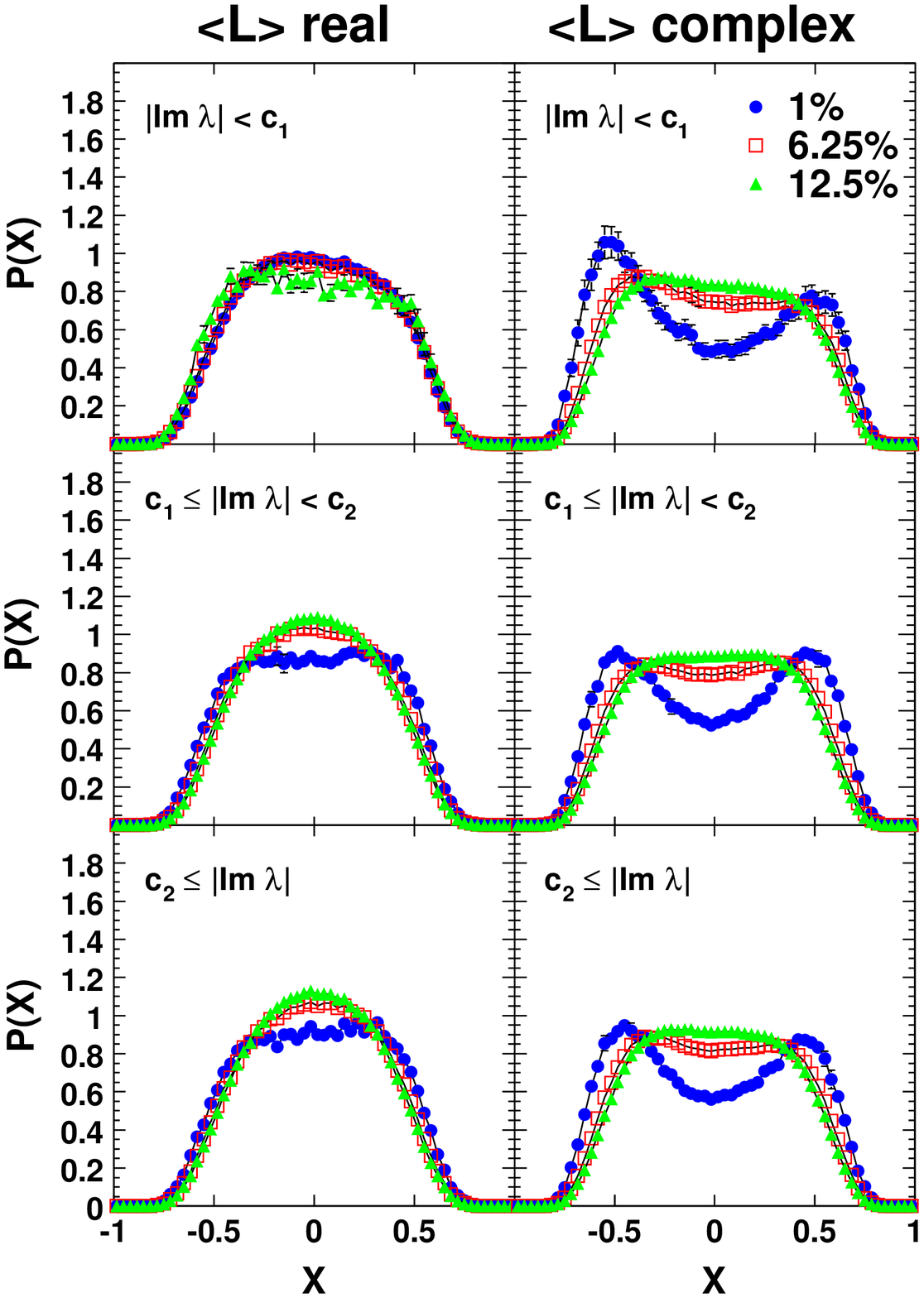,height=14.5cm,clip} 
\caption{Local chirality for non zero modes in the chirally symmetric phase
($\beta_1 = 8.45, 6\times 20^3$). We divide our data with respect to
the sector of the Polyakov loop. The left column of plots 
contains the results 
for real $\langle L \rangle$, the right column shows data for complex
$\langle L \rangle$. In addition we bin the data
with respect to $| \mbox{Im } \lambda|$ such that the first bin
(top plots) shows the local chirality for eigenvectors with 
eigenvalues close to the edge of the spectral gap, while
the other bins (middle and bottom plots) show eigenvectors with 
eigenvalues more remote from the edge.
\label{lchir845}}
\end{center}
\end{figure}

Let us now turn to the non zero modes in the chirally broken phase
($\beta_1 = 8.10$). In Fig.~\ref{lchir810} we show our results. Again they
were computed from 55 gauge configurations on the $6 \times 20^3$ lattice
and we use the same cut-off values as for the zero modes, i.e.~1\%,
6.25\% and 12.5\%. In addition we binned the eigenvectors used for 
the computation of the local chirality with respect to the imaginary
parts of the corresponding eigenvectors. We divided all eigenvectors
into three bins such that their eigenvalues obey
$| \mbox{Im } \lambda | < c_1$ for the first bin,  
$c_1 \leq | \mbox{Im } \lambda | < c_2$ for the second bin and  
$c_2 \leq | \mbox{Im } \lambda |$ for the third bin. 
The thresholds $c_i$ were set such that on the average the
first bin contained 4 eigenvectors and the other two bins each 8
eigenvectors (for the computation of the local chirality we always 
stored a total of 20 eigenvectors which are not zero modes).
The values of the $c_i$ are given by $c_1 = 0.018$ and $c_2 = 0.057$.
The first bin (top plot in Fig.~\ref{lchir810})
contains the very near zero modes, while in the
second and third bin (middle and bottom plots in  Fig.~\ref{lchir810})
we give the results for eigenvectors with larger $| \mbox{Im } \lambda |$.

For the very near zero modes ($| \mbox{Im } \lambda | < c_1$, top plot
in Fig.~\ref{lchir810}) we find a very pronounced signal for
the local chirality. In fact it is almost indistinguishable from
the result for the zero modes shown in the top plot
of Fig.~\ref{lchirzero} (note the different scale on the 
vertical axis in Fig.~\ref{lchirzero} and Fig.~\ref{lchir810}).
This shows that the topological molecules responsible
for the near zero modes are still very much 
reminiscent of the localized structure
of the two partners, i.e.~the caloron and the anti-caloron.
Furthermore the dependence for the cut-off 
on the number of supporting points is very weak, i.e.~the 1\% and
the 12.5\% curves are still very similar. This indicates that these
modes are relatively spread out, such that, similar to the zero modes,
more than one eighth of the volume is occupied by excitations
with a definite chirality.

As the eigenvalues move away from the origin, the local chirality
of the corresponding eigenvectors becomes less pronounced. This can be seen 
from the middle and bottom plots in Fig.~\ref{lchir810} where we show
our results for the second and third bin of   
$| \mbox{Im } \lambda |$. Firstly we find that the peaks are less 
pronounced and secondly, that now the curve is much more sensitive
to the cut-off for the percentage of supporting lattice points.
The interpretation of both these observations is straightforward.
According to the instanton picture, eigenvalues higher up in the spectrum 
are produced by pairs of topological excitations which are overlapping
more than the excitations for the near zero modes. Thus the localization
for the two partners in a molecule is more washed out and so is
the local chirality. This washing out is also reflected in the
stronger dependence on the percentage for the supporting lattice points. 
If the overall height of the peaks in $p(x)$ is going down, then the
1\% cut may still see only the highest peaks which are still locally 
chiral, but the larger values 6.25\% and 12.5\% start to cut also
into the quantum fluctuations which do not have local chirality.
To summarize the discussion of Fig.~\ref{lchir810}, we find good agreement
with the picture of topological molecules which produce the
nonvanishing eigenvalue density near the origin and lead to eigenvalues
higher up (down) in the spectrum as the two partners in the molecule
start to overlap more.

Finally we discuss our results for the non zero modes in the chirally
symmetric phase ($\beta_1 = 8.45, 6\times20^3$). 
Fig.~\ref{lchir845} shows our data which were again divided into
the two sectors for the Polyakov loop and the left column of plots 
shows the sector with $\langle L \rangle$ real and the 
right column shows the complex sector. As in the chirally broken phase we 
bin the eigenvectors with respect to the imaginary parts of the
corresponding eigenvalues. The first bin 
($| \mbox{Im } \lambda | < c_1$, top plot in Fig.~\ref{lchir845})
contains the data for eigenvectors with eigenvalues very 
close to the edge of the gap, while the
other two bins ($c_1 \leq | \mbox{Im } \lambda | < c_2$, middle plot
and $c_2 \leq | \mbox{Im } \lambda |$, bottom plot) 
show the results for eigenvectors with eigenvalues more remote from 
the edge. Again we chose the values of the thresholds $c_i$ such 
that the first bin contains an average of 4 eigenvectors while the
other two bins each contain 8 eigenvectors. This binning is
done for each of the sectors individually. The resulting values for the
thresholds at $\beta_1 = 8.45$ are then given by 
$c_1 = 0.207$ and $c_2 = 0.230$ for the real sector and
$c_1 = 0.081$ and $c_2 = 0.112$ for the complex sector.

It is obvious, that for the real sector (left column of plots) 
the signal for local chirality has vanished entirely. For all three 
bins and for all three percentages of supporting lattice points
we do not see a dip in $P(X)$. This shows that eigenvectors
with local chirality do not play a
substantial role in the real sector of the chirally symmetric phase at 
$\beta_1 = 8.45$. Similarly in the data for the complex sector of the
Polyakov loop only the 1\% cut-off, i.e.~only the highest peaks of the 
scalar density, produces a small dip in $P(X)$, which, however, vanishes 
as one increases the percentage of supporting points to 
6.25\% or even 12.5\%. Thus only about 1\% of the fluctuations 
in the complex sector of the $\beta_1 = 8.45$ data show local chirality.
Based on Fig.~\ref{i2scal} and the subsequent discussion we
interpret the failure to find non-zero modes with local chirality
as further evidence for the diluteness of topological molecules
in the chirally symmetric phase.

\section{Summary}

In this article we have presented a large scale study of eigenvectors
and eigenvalues of the lattice Dirac operator. 
We analyzed the Dirac operator for
background gauge fields from quenched finite temperature ensembles on
both sides of the QCD phase transition. Using three different lattice sizes
we analyzed the volume dependence of our observables. We used 
a Dirac operator which is an approximate solution of the
Ginsparg-Wilson equation and has good chiral properties. 

Our observables focused on testing the topological nature of the gauge field
excitations as seen by the Dirac operator. In particular we analyzed
the density of the eigenvalues, the topological charge
defined through the index theorem, scalar and pseudoscalar densities for
the eigenvectors, their localization properties and a 
recently proposed observable for the local chirality of the
eigenvectors. Our main results are:

\begin{itemize}

\item 

For the low temperature phase we find a density $\rho(\lambda)$ 
of eigenvalues which extends all the way to the origin which, according to 
the Banks Casher relation, gives rise to a non-vanishing chiral condensate 
in this phase. In the high temperature phase we observe a gap in the
spectral density and thus a restoration of chiral symmetry.

\item 

The spectral gap opens up for all phases of the Polyakov loop, i.e.~when 
we divide our ensemble into configurations with real Polyakov loop 
and configurations where the Polyakov loop has a phase of $\pm 2 \pi/3$
we find a gap for both subsets. This shows that the chiral symmetry 
is restored in both these sectors.

\item

When analyzing the topological charge $Q$ as defined by the index theorem
we find a probability distribution and values of $\langle Q^2 \rangle$ which
can be described  well by a simple model of topological excitations.
The drop of $\langle Q^2 \rangle$ as one increases $\beta_1$ 
reflects a thinning out of the topological excitations
which prevents the formation of a chiral condensate in the high temperature 
phase.

\item

Plots of the pseudoscalar densities for a zero mode and a near zero mode  
illustrate the characteristic behavior of the different types of excitations.

\item

A measure for 
the localization of the eigenvectors is given by the so called 
inverse participation ratio. For the low temperature
phase, we found that the most localized
eigenvectors have eigenvalues
near the origin and the localization of eigenvectors gets washed out as 
the imaginary part of the corresponding eigenvalue increases. Similarly
one finds for the high temperature phase 
that the most localized eigenvectors have eigenvalues near the
edge of the distribution. These observations support the picture 
that near zero modes, respectively the near edge modes,  
are the still relatively localized remnants of 
interacting calorons and anti-calorons.

\item

  In the high temperature phase we find that the eigenmodes in the
 {\bf Z}$_3$ sector where the Polyakov loop is real are 
 more strongly localized than modes in the complex {\bf Z}$_3$ sector.
 This behavior supports the idea that calorons are relevant, 
 because the classical solution of the Dirac equation in a caloron  
 gauge field shows the same behavior, see Appendix B. 

\item

When plotting the number of eigenvectors with a large
localization as a function of $\beta_1$ we found a drop when 
increasing $\beta_1$, similar to the 
drop observed for $\langle Q^2 \rangle$. This shows that also interacting 
pairs of calorons and anti-calorons become less abundant at larger 
$\beta_1$ and in the high temperature phase their number is
no longer sufficient to build up a non-vanishing chiral condensate.

\item

Finally we analyzed the local chirality of the eigenvectors at different
values of the cut-off for the scalar density. For the chirally
broken phase our results for the near zero modes 
confirm the pattern of local chirality as
expected for molecules of topological excitations. In the chirally symmetric
phase we still see modes with local chirality near the edge of the gap,
but they are much more diluted by bulk modes dominated by quantum 
fluctuations.

\end{itemize}

Our results strongly support a picture of interacting 
topological excitations which are responsible for the breaking of
the chiral symmetry at low temperatures, while in the high temperature
phase they are no longer sufficiently abundant and the chiral
symmetry is restored.
\\
\\
{\bf Acknowledgements: } We would like to thank 
Andrei Belitsky, Peter Hasenfratz, Holger Hehl,
Ivan Hip, Tamas Kovacs, Christian B.~Lang, Ferenc Niedermayer,
Edward Shuryak, Wolfgang S\"oldner and Christian Weiss for
interesting discussions. This project was supported by 
the Austrian Academy of Sciences, the DFG and the BMBF. We thank 
the Leibniz Rechenzentrum in Munich for computer time on the Hitachi 
SR8000 and their operating team for training and support.

\begin{appendix}
\setcounter{equation}{0}
\section{Detailed specification of the chirally improved Dirac operator}
In this appendix we describe in more detail the terms in our Dirac operator 
and give the values for the coefficients which we use for the 
four ensembles of quenched gauge field configurations. 

\begin{table}[h]
\begin{center}
\begin{tabular}{ c c c }
\hline
 & & \\
Clifford generator & Generating path & Name of coefficient \\
 & & \\
\hline
1\hspace{-1.0mm}I & $ < > $ & $s_1$ \\
1\hspace{-1.0mm}I & $ <1> $ & $s_2$ \\
1\hspace{-1.0mm}I & $ <1,2 > $ & $s_3$ \\
1\hspace{-1.0mm}I & $ <1,2,3 > $ & $s_5$ \\
1\hspace{-1.0mm}I & $ <1,1,2 > $ & $s_6$ \\
1\hspace{-1.0mm}I & $ <1,2,-1 > $ & $s_8$ \\
1\hspace{-1.0mm}I & $ <1,2,3,4 > $ & $s_{10}$ \\
1\hspace{-1.0mm}I & $ <1,2,-1,3 > $ & $s_{11}$ \\
1\hspace{-1.0mm}I & $ <1,2,-1,-2 > $ & $s_{13}$ \\
\hline
$ \gamma_1$ & $ <1> $ & $v_1$ \\
$ \gamma_1$ & $ <1,2> $ & $v_2$ \\
$ \gamma_1$ & $ <1,2,3> $ & $v_4$ \\
$ \gamma_1$ & $ <2,1,3> $ & $v_5$ \\
\hline
$ \gamma_1 \gamma_2 $ & $ <1,2> $ & $t_1$ \\
$ \gamma_1 \gamma_2 $ & $ <1,2,3> $ & $t_2$ \\
$ \gamma_1 \gamma_2 $ & $ <1,3,2> $ & $t_3$ \\
$ \gamma_1 \gamma_2 $ & $ <1,2,-1> $ & $t_5$ \\
$ \gamma_1 \gamma_2 $ & $ <1,2,-1,-2> $ & $t_{15}$ \\
\hline
$ \gamma_5 $ & $ <1,2,3,4> $ & $p_1$ \\
\hline
\end{tabular}
\end{center}
\caption{Description of the terms in our $D$.}
\label{terms}
\end{table}

As has been pointed out in  Section 2.1, the most general Dirac operator $D$
can be expanded in the  series (\ref{gendirac}). Each term in this series is
characterized by three  pieces: A generator of the 
Clifford algebra, a group of paths and a real coefficient. 
The paths within a group can have different signs which are determined by the
symmetries, C, P,  $\gamma_5$-hermiticity and rotation invariance.
The symmetries also determine
which paths are grouped together. Thus it is sufficient to characterize a group
of paths by a single generating path and all the other paths in the group as
well as their relative sign factors can be  determined by applying the
symmetries. 

In addition, for the vector and tensor terms appearing in our $D$ it is 
sufficient to give the paths only for one vector (tensor) since rotation
invariance immediately fixes the structure for the other vector (tensor) 
terms. In order to describe our $D$ we start with listing the three determining
pieces for each term in Table~\ref{terms}. 

In Table~\ref{coeffvals} we list the values of the coefficients
$s_i,v_i,t_i$ and $p_1$
for the different values of the inverse gauge coupling
$\beta_1$ for the ensembles used in this work. 

\begin{table}[h]
\begin{center}
\begin{tabular}{ c c c c c }
\hline
 & $\beta_1 = 8.10$ & $\beta_1 = 8.20$ & 
$\beta_1 = 8.30$ & $\beta_1 = 8.45$ \\
\hline
 $s_1$  & $+1.54498$ & $+1.54590$ & $+1.54688$ & $+1.54737$ \\
 $s_2$  & $-0.06169$ & $-0.06063$ & $-0.05997$ & $-0.05892$ \\
 $s_3$  & $-0.01448$ & $-0.01449$ & $-0.01448$ & $-0.01451$ \\
 $s_5$  & $-0.00262$ & $-0.00258$ & $-0.00255$ & $-0.00251$ \\
 $s_6$  & $+0.00220$ & $+0.00215$ & $+0.00210$ & $+0.00206$ \\
 $s_8$  & $-0.00540$ & $-0.00534$ & $-0.00532$ & $-0.00525$ \\
 $s_{10}$ & $-0.00053$ & $-0.00052$ & $-0.00051$ & $-0.00050$ \\
 $s_{11}$ & $-0.00118$ & $-0.00117$ & $-0.00117$ & $-0.00116$ \\
 $s_{13}$ & $+0.00780$ & $+0.00778$ & $+0.00777$ & $+0.00775$ \\
\hline
 $v_1$  & $+0.10975$ & $+0.11083$ & $+0.11063$ & $+0.11240$ \\
 $v_2$  & $+0.01770$ & $+0.01724$ & $+0.01707$ & $+0.01654$ \\
 $v_4$  & $+0.00744$ & $+0.00767$ & $+0.00784$ & $+0.00804$ \\
 $v_5$  & $+0.00182$ & $+0.00193$ & $+0.00198$ & $+0.00209$ \\
\hline
 $t_1$  & $-0.09874$ & $-0.09862$ & $-0.09860$ & $-0.09843$ \\
 $t_2$  & $-0.00309$ & $-0.00303$ & $-0.00299$ & $-0.00292$ \\
 $t_3$  & $+0.00227$ & $+0.00224$ & $+0.00221$ & $+0.00218$ \\
 $t_5$  & $-0.00663$ & $-0.00651$ & $-0.00641$ & $-0.00627$ \\
 $t_{15}$ & $-0.00361$ & $-0.00358$ & $-0.00356$ & $-0.00354$ \\
\hline
 $p_1$  & $-0.00911$ & $-0.00907$ & $-0.00905$ & $-0.00901$ \\
\hline  
\end{tabular}
\end{center}
\caption{The numerical values of the 
coefficients $s_i, v_i$, $t_i$ and $p_1$ for
the different ensembles of gauge fields.
\label{coeffvals}}
\end{table}

\newpage
\setcounter{equation}{0}

\section{Quark zero modes}

 \subsection{Solutions of the field equation}

 In this section we discuss the quark zero modes  
 in the caloron gauge field
 solutions of Harrington and Shepard~\cite{HS}. 
 In particular, we want to investigate their dependence 
 on the {\bf Z}$_3$ sector. 

  't Hooft has shown that one way of generating multi-instanton
 solutions of the gluon equations of motion is to consider
 gauge fields of the form
 \begin{equation}
 A^\mu_{c c^\prime}(x) = i {\bar \sigma}^{\mu \nu}_{c c^\prime}
        \; \partial_\nu \ln \rho(x) \; ,
 \label{thooft}
 \end{equation}  
 where $c$ and $c^\prime$ are color indices and 
 the ${\bar \sigma}$ are a set of $2\times 2$ color matrices. 
 A suitable basis for the ${\bar \sigma}$ is given in~\cite{jackiw}. 
 After a little algebra one finds that this $A$ field satisfies
 the equations of motion if 
 \begin{eqnarray} 
 &&\partial^2 \rho(x) = 0 \label{boxrho} \\
{\rm or}\quad && \partial^2 \rho(x) \propto \rho^3(x) \,.
 \end{eqnarray} 
 We know from the Atiyah-Singer index theorem that gauge field 
 configurations with topological charge must have fermion zero
 modes. Following~\cite{jackiw} these modes can be written in a
 form analogous to (\ref{thooft}), 
 \begin{equation}
 \psi_{c d}(x) = \rho^{1/2}(x) \; \alpha^\mu_{c d} 
 \, \partial_\mu \left( \frac{ \chi(x)}{\rho(x)} \right) \; ,
 \label{psiansatz}
 \end{equation} 
 where $c$ is a color index, and $d$ a Dirac index. 
 The relation of the $\alpha$ matrices to the ${\bar \sigma}$
 matrices, and a choice of $\alpha$ is given in~\cite{jackiw}. 
 The ansatz (\ref{psiansatz}) gives a zero mode of the Dirac
 operator if 
 \begin{equation}
 \rho(x) \, \partial^2 \chi(x) - \chi(x) \, \partial^2 \rho(x) = 0 \,. 
 \label{boxchi} 
 \end{equation} 

  We now need a solution of Eq.~(\ref{boxrho}) that also satisfies
 the periodic boundary condition in the time direction, 
 \begin{equation} 
  \rho(\vec{r}, t + 1/T) = \rho(\vec{r}, t) \; ,
 \end{equation} 
 where $T$ is the temperature (we have set Boltzmann's constant 
 to 1). A simple solution of this type, \cite{HS}, is 
 \begin{equation} 
  \rho(\vec{r},t) =
   1 + \sum_{k=-\infty}^{\infty} \frac{R^2}{r^2 + (t-k/T)^2} 
  = 1 + \, \frac{\pi T R^2 \; \sinh(2 \pi T r)}
  { r \left[ \cosh(2 \pi T r)- \cos(2 \pi T t)\right] } 
  \label{rhocal} 
 \end{equation} 
 where $r^2 \equiv r_1^2 + r_2^2 + r_3^2$. 
 This solution corresponds to a caloron of radius $R$ sitting
 at the origin. We can use translation invariance to give 
 solutions at other locations, and add together solutions of the above
 form to get multi-caloron solutions. 

 We now look for a fermion zero mode in the solution (\ref{rhocal}), 
 which should obey anti-periodic boundary conditions
 \begin{equation} 
  \psi(\vec{r}, t + 1/T) = -\psi(\vec{r}, t) \;. 
 \end{equation} 
 $\psi$ depends linearly on $\chi$, so $\chi$ must obey the 
 same boundary conditions as  $\psi$. The function 
\begin{eqnarray}
\chi(\vec{r},t) &=& \frac{1}{2 \pi}
\sum_{k=-\infty}^{\infty} (-1)^k \; \frac{R}{r^2 + (t-k/T)^2} 
 \nonumber \\
 &=& \frac{ T R \; \cos(\pi T t) \; \sinh( \pi T r)} 
{ r \left[ \cosh(2 \pi T r)- \cos(2 \pi T t)\right] } \; ,
\label{reZ}
\end{eqnarray} 
satisfies eq.~(\ref{boxchi}), and has the correct behavior 
under $t \to t +1/T$. This has the same form as the general 
multi-instanton solution given in \cite{solve}.

The solution (\ref{reZ}) corresponds to the real sector of 
the {\bf Z}$_3$ symmetry. We are also interested in knowing what 
the zero-mode looks like in the complex sectors. There are 
two possible approaches to this problem. 

The direct approach is to use the fact that eq.~(\ref{thooft})
gives an SU(2) $A$-field. When we embed this solution in SU(3)
the $A$-field has components proportional to the three generators
$t^1, t^2,$ and $t^3$ of SU(3). The generator $t^8$ commutes 
with these three matrices, so we still have a solution of the 
equation of motion if we add an arbitrary constant $A$ field
in the 8-direction of SU(3). This can be used to produce a caloron 
solution with an arbitrary Polyakov loop value. See \cite{vBaal}
for methods of constructing SU(2) solutions with non-trivial 
Polyakov loop.
 
\begin{figure}[p]
\begin{center} 
\epsfig{file=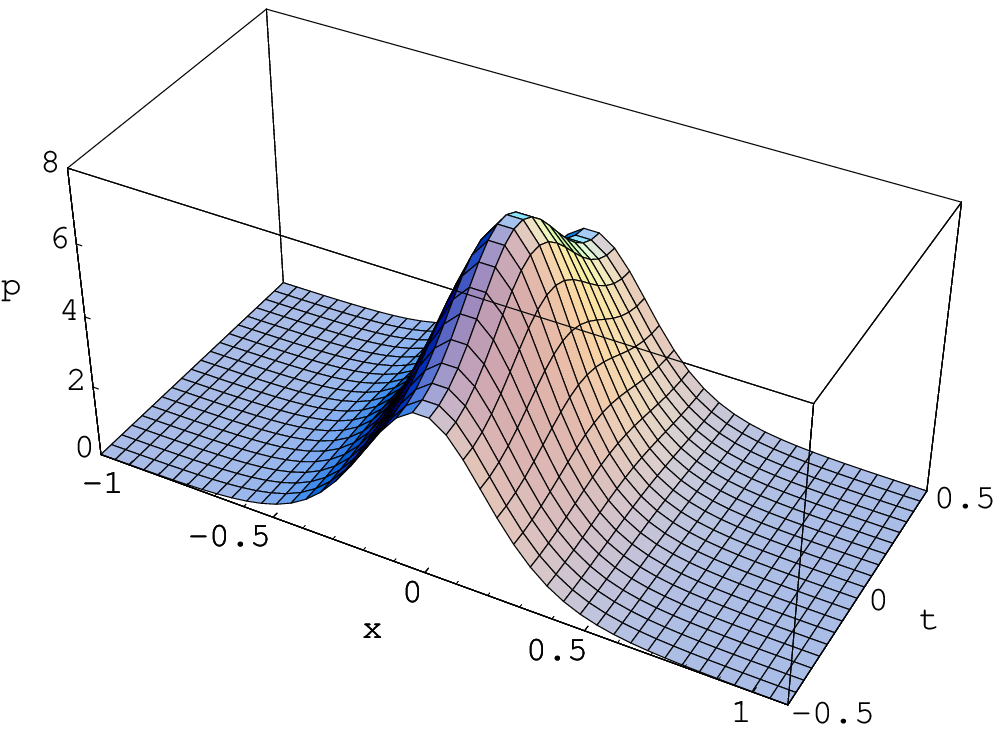, width = 10cm}
\end{center}
\begin{center}
\epsfig{file=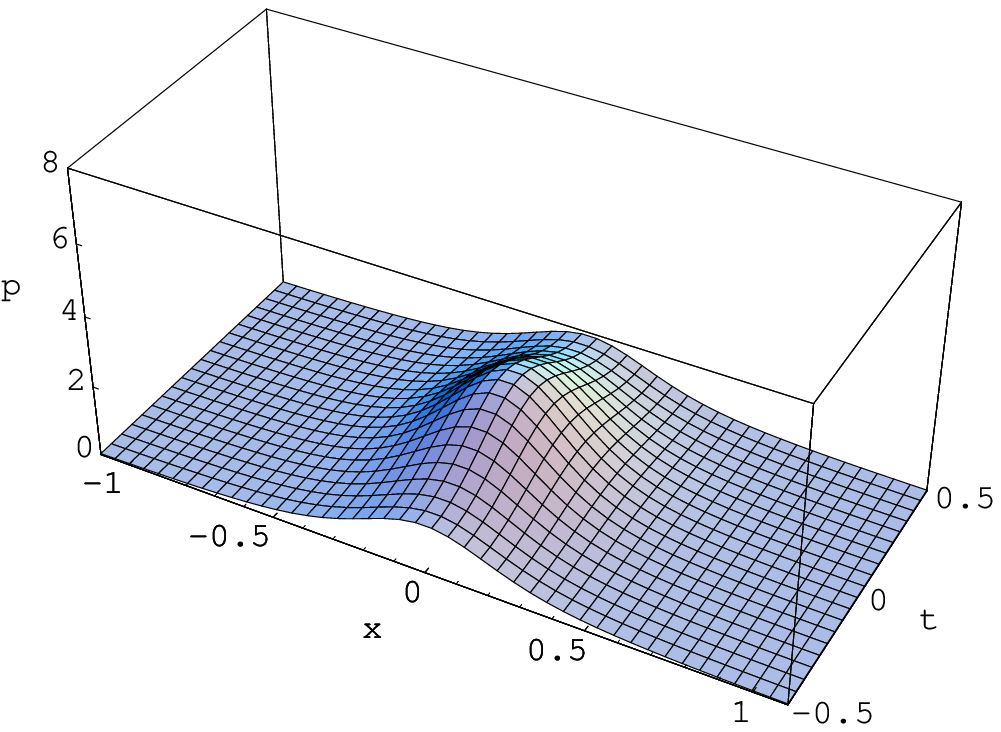, width = 10cm}
\end{center}

\caption{The scalar density $p(x)$ for a quark zero mode 
in the real {\bf Z}$_3$ sector (upper plot)
and the complex {\bf Z}$_3$ sector (lower plot). 
In both cases the caloron radius is $1/T$. }
\label{mathmode} 
\end{figure} 

A simpler method is to note that a Polyakov loop can be 
gauge transformed to look just like a fermion boundary 
condition of the form  
\begin{equation} 
\psi(\vec{r}, t + 1/T) = - e^{ i\theta} \psi(\vec{r}, t)\,.
\label{theta_bc}
\end{equation} 
For the complex {\bf Z}$_3$ sectors of high-temperature QCD we are 
interested in $\theta = \pm 2 \pi/3$.  A solution of
eq.~(\ref{boxchi}) that obeys Eq.~(\ref{theta_bc}) is given by
\begin{eqnarray} 
\lefteqn{\chi(\vec{r},t) = \frac{1}{2 \pi}
\sum_{k=-\infty}^{\infty} (-1)^k e^{i k \theta}
\; \frac{R}{r^2 + (t-k/T)^2} }
\label{compZ}
\\
&=& \frac{ e^{i\theta T t} 
TR \left[ \cosh(\theta T r) \cos(\pi T t) \sinh( \pi T r)
-i \sinh(\theta T r) \sin(\pi T t) \cosh(\pi T r)\right] } 
{ r \left[ \cosh(2 \pi T r)- \cos(2 \pi T t)\right] } , 
\nonumber 
\end{eqnarray} 
valid for $-\pi \le \theta \le \pi$. 
At large $r$ the probability density found from eq.~(\ref{compZ})
drops off like $r^{-2} \exp[ -2 (\pi - |\theta|) T r]$. This
is much faster in the real sector ($\theta=0$) than in the 
complex sectors ($\theta = \pm 2\pi/3$). 

\subsection{Phenomenological implications}

To illustrate the difference between the {\bf Z}$_3$ sectors,
we plot the density of the fermion zero mode in a 
caloron of radius $R = 1/T$, see Fig.~\ref{mathmode}.
We show a two-dimensional
$x$-$t$ slice through the center of the mode. Both 
modes have been normalized so that the 4-dimensional
integral of $p$ is $1$. One sees clearly that the 
mode in the real sector is more strongly localized.     

\begin{figure}[h] 
\begin{center} 
\epsfig{file=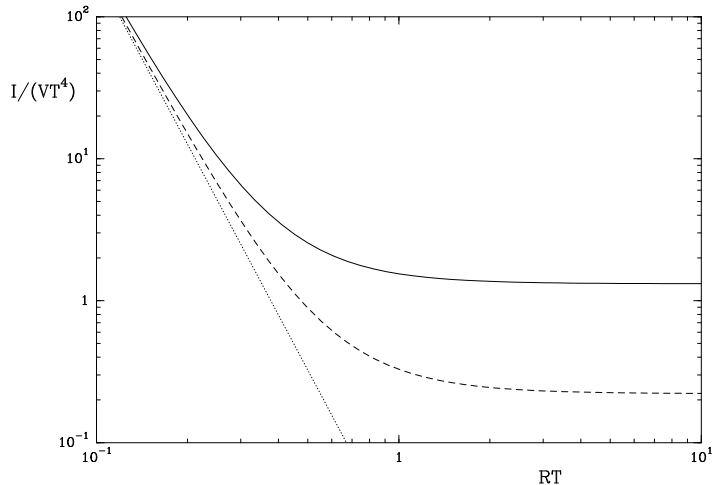, angle =270, width = 10cm}
\end{center}
\caption{The inverse participation ratio $I$ of the quark zero-mode
calculated for a caloron in the real {\bf Z}$_3$ (solid line)
and in the complex sector (dashed line). For comparison the result
for an instanton is shown as the dotted line. $R$ is the radius of the
caloron and $T$ the temperature.}
\label{Itheory} 
\end{figure} 

A systematic way to compare the localization of the 
modes  (\ref{reZ}) and  (\ref{compZ}) is to define an
inverse participation ratio analogous to that of Section 4. 
The scalar density $p(x)$ is defined in exactly the same way
in the continuum as on the lattice, Eq.~(\ref{densdef}).
 In the definition of $I$ we simply replace the sum
 in Eq.~(\ref{iprdef}) by an integral: 
\begin{equation} 
I/V \equiv \int d^4x \; p(x)^2 \;
\end{equation} 
where the eigenmode has been normalized so that 
$\int d^4x \; p(x) =1$. 

In Fig.~\ref{Itheory} we compare the
localization of the quark zero modes
as the caloron radius $R$ is varied. When $R T \ll 1$ thermal 
effects are small, so a caloron looks very like an instanton,
and the $I$ of the zero mode has only a slight dependence on
the {\bf Z}$_3$ sector. This contrasts with the situation when 
the caloron radius is comparable with the inverse temperature, 
$R T \sim 1$. Here the {\bf Z}$_3$ sector has a dramatic effect, 
with localization being much stronger in the real sector
than in the complex. This is reminiscent of the effects 
we see in the high temperature phase (see Section 4). 

\end{appendix}

\end{document}